\documentclass[twocolumn,trackchanges]{aastex62}
\usepackage{natbib}
\usepackage{xspace}
\newcommand{\icnu}{IceCube-170922A\xspace}
\newcommand{\txs}{TXS\,0506+056\xspace}
\newcommand{\RN}[1]{%
  \textup{\uppercase\expandafter{\romannumeral#1}}%
}

\begin{document}

\title[The neutral beam model for TXS 0506+056]{A Neutral Beam Model for High-Energy Neutrino Emission from the Blazar TXS 0506+056}

\author{B. Theodore Zhang}
\affiliation{Department of Physics, The Pennsylvania State University, University Park, Pennsylvania 16802, USA}
\affiliation{Department of Astronomy \& Astrophysics, The Pennsylvania State University, University Park, Pennsylvania 16802, USA}
\affiliation{Center for Particle and Gravitational Astrophysics, The Pennsylvania State University, University Park, Pennsylvania 16802, USA}
\affiliation{Department of Astronomy, School of Physics, Peking University, Beijing 100871, China }
\affiliation{Kavli Institute for Astronomy and Astrophysics, Peking University, Beijing 100871, China}
\author{Maria Petropoulou}
\affiliation{Department of Astrophysical Sciences, Princeton University, New Jersey 08544, USA}
\author{Kohta Murase}
\affiliation{Department of Physics, The Pennsylvania State University, University Park, Pennsylvania 16802, USA}
\affiliation{Department of Astronomy \& Astrophysics, The Pennsylvania State University, University Park, Pennsylvania 16802, USA}
\affiliation{Center for Particle and Gravitational Astrophysics, The Pennsylvania State University, University Park, Pennsylvania 16802, USA}
\affiliation{Yukawa Institute for Theoretical Physics, Kyoto, Kyoto 606-8502, Japan}
\author{Foteini Oikonomou}
\affiliation{European Southern Observatory, Karl-Schwarzschild-Str. 2, Garching bei M{\"u}nchen D-85748, Germany}

\begin{abstract}
The IceCube collaboration reported a $\sim 3.5\sigma$ excess of $13\pm5$ neutrino events in the direction of the blazar \txs during a $\sim$6 month period in 2014--2015, as well as the ($\sim3\sigma$) detection of a high-energy muon neutrino during an electromagnetic flare in 2017.
We explore the possibility that the 2014--2015 neutrino excess and the 2017 multi-messenger flare are both explained in a common physical framework that relies on the emergence of a relativistic neutral beam in the blazar jet due to interactions of accelerated cosmic rays (CRs) with photons. We demonstrate that the neutral beam model provides an explanation for the 2014--2015 neutrino excess without violating X-ray and $\gamma$-ray constraints, and also yields results consistent with the detection of one high-energy neutrino during the 2017 flare. If both neutrino associations with TXS 05065+056 are real, our model requires that (i) the composition of accelerated CRs is light, with a ratio of helium nuclei to protons $\gtrsim5$, (ii) a luminous external photon field ($\sim 10^{46}$~erg s$^{-1}$) variable (on year-long timescales) is present, and (iii) the CR injection luminosity as well as the properties of the dissipation region (i.e., Lorentz factor, magnetic field, and size) vary on year-long timescales.
\end{abstract}


\section{Introduction} \label{sec:intro} 
The IceCube Collaboration recently reported the detection of a high-energy ($E_\nu\gtrsim290$ TeV) muon-track neutrino event (\icnu) from the flaring blazar \txs~\citep{IceCube:2018dnn}, providing the $\sim3\sigma$ high-energy neutrino-source association. A follow-up analysis of IceCube archival data revealed a past ``neutrino excess'' at a significance level of $\sim 3.5\sigma$ ($13\pm5$ signal events within $\sim$ 6 months) from the direction of \txs, which, however, was not accompanied by an electromagnetic flare \citep{IceCube:2018cha, Garrappa2019}. The most probable energy for the neutrinos from the 2014--2015 period lies in the range $\sim 10 - 100\rm~TeV$ and the inferred isotropic muon neutrino luminosity, if all signal events originated from \txs, is $\simeq 1.2 \times 10^{47}\rm~erg~s^{-1}$~\citep{IceCube:2018cha}.

Blazars are active galactic nuclei (AGN) with strong relativistic jets oriented at a small angle with respect to the line of sight~\citep[e.g.,][]{Urry1995}. Being some of the most powerful astrophysical steady sources, blazars have been extensively studied as sources of high-energy astrophysical cosmic rays (CRs) and neutrinos~\cite[see, e.g.,][]{Mannheim:1992, Mannheim:1995, 1997ApJ...488..669H, Atoyan:2002gu,Murase:2014foa,Dermer:2014vaa,Petropoulou2015,Rodrigues:2017fmu, Palladino:2018lov,Oikonomou:2019djc}.

Theoretically, flares are ideal periods of neutrino production in blazars. During flaring episodes, the density of the target photon field for photomeson interactions with the hadrons in the blazar jet is usually enhanced. It is also possible that the injection rate of accelerated protons is  simultaneously enhanced. As a result, many models predict that the neutrino luminosity, $L_{\nu}$, is strongly enhanced during flares, with $L_{\nu} \propto L_{\gamma}^{\alpha}$, where $L_{\gamma}$ is the photon luminosity and $\alpha \sim 1.5 - 2$~\citep{Murase:2014foa,Tavecchio:2014iza, Petropoulou:2016ujj,Murase:2016gly}. From the experimental point of view, flares constitute ideal periods for neutrino emission, as the rate of background (atmospheric) neutrinos is reduced by focusing searches on a narrow time window. 

The reported association of \icnu with the 2017 flare of \txs was studied in detail by several authors in the context of scenarios invoking photo-hadronic interactions \citep[e.g.,][]{Keivani:2018rnh,Murase:2018iyl,Gao:2019NatAs,Cerruti:2018tmc} or hadro-nuclear collisions \citep[e.g.,][]{Murase:2018iyl,Liu2019,Sahakyan2018} for neutrino production. Most of the aforementioned studies concluded that at most $\sim 0.01-0.1$ muon neutrinos\footnote{At most $\sim0.01$ muon neutrinos could have been detected through the EHE alert channel within six months in the modeling that took into account the ultraviolet data~\citep{Keivani:2018rnh}.} could have been produced by \txs during the $\sim6$-month-long electromagnetic flare, if the neutrino emission originated from the same region in the blazar as the bulk of the photon emission (single-zone scenarios). Slightly higher  neutrino production rates can be obtained (by a factor of $\sim 10$), if the production sites of neutrinos and low-energy photons are decoupled \citep[see e.g.,][]{Murase:2018iyl,Xue2019}. Nevertheless, the observation of \textit{one} high-energy neutrino from the 2017 flare of \txs is consistent with the theoretical estimates in the presence of an ensemble of faint sources with summed expectation $\sim$one neutrino~\citep{IceCube:2018dnn,Strotjohann:2018ufz}, but it implies that this association was lucky. 

The lack of an electromagnetic flare during the 2014--2015 neutrino excess does not fit well in the above picture\footnote{It has been also proposed that the observed 2014--2015 neutrino excess has contributions not only from \txs, but also from the nearby FSRQ PKS~0502+049~\citep{Liang:2018siw, Banik:2019twt}. However, there is no consensus whether the 2014--2015 neutrino excess could originate in PKS~0502+049~\citep[see e.g.][]{Padovani:2018acg}.}
\cite{Murase:2018iyl} pointed out that single-zone scenarios lead to cascade X-ray emission detectable by the X-ray satellites {\it Swift} and {\it MAXI}, assuming that jet parameters are similar to those of the 2017 flare. 
Studies by \cite{Reimer:2018vvw} and \cite{Rodrigues:2018tku} found no parameters that can explain the 2014--2015 neutrino excess in a single-zone model. Petropoulou et al. (in preparation) also found no single-zone model -- among the fifty models that they explored -- that can explain the neutrino flux and simultaneously satisfy the electromagnetic constraints from the spectral energy distribution (SED) of \txs. 

The aforementioned results highlight the need for multi-zone models to explain the 2014--2015 excess of neutrinos in the direction of \txs, 
In practice, the existence of more than one emitting regions in the jet of \txs means the possibility to balance the relative energy output of the different emitting species across different parts of the jet, albeit at the cost of additional free parameters. Here, we explore one such model, based on the idea that accelerated cosmic-ray  nuclei in the inner jet interact with internal synchrotron photons and external radiation fields to produce a collimated beam of neutrons, in addition to neutrinos and $\gamma$-rays~\citep{Eichler:1978,Atoyan:2002gu,Dermer:2012rg}. The neutrons continue to interact with external photon fields on parsec (pc) scales and produce additional neutrinos, which can dominate the total neutrino output of the blazar. In contrast to single-zone models previously reviewed, the electromagnetic cascade induced by the beam can be suppressed in this setup due to (i) the isotropization and time delay of electrons and positrons in the large scale jet and (ii) the lack of pair injection due to the Bethe-Heitler (BH) process which is irrelevant for neutrons.

\citet{Murase:2018iyl} have shown with analytical arguments that the {\it neutral beam} model can, in principle, explain the 2014--2015 and 2017 neutrino emission of \txs. In this work, we perform a detailed numerical investigation of the neutral beam model with the goal of explaining the multi-messenger observations of \txs in 2014--2015. For this purpose, we numerically compute the electromagnetic and neutrino emissions produced in the inner emitting blob and the neutral beam for a wide set of parameters. We also explore if the same framework can be applied to the 2017 neutrino detection of \icnu.

This paper is organized as follows. 
In Section~\ref{sec:2}, we outline the neutral beam model and present analytical estimates for the production efficiency of neutrinos and secondary electron-positron pairs by the interactions of nuclei, protons, and neutrons with photons.  
In Section~\ref{sec:3}, we present the numerical approach we adopted for calculating the neutrino and photon emission within our model.  In Section~\ref{sec:4}, we present numerical results on the expected neutrino flux and accompanying electromagnetic emission from the blazar \txs \, in 2014--2015 and 2017. In Section~\ref{sec:5}, we discuss the implications of our results and conclude with a summary.

\section{Model description}\label{sec:2}
\subsection{General considerations}
Blazars -- active galactic nuclei with relativistic jets pointing towards the observer \citep[e.g.,][]{1993ARA&A..31..473A,Urry1995} -- are thought to be powered by an accreting supermassive black hole (SMBH) in their centers
\citep[e.g.,][]{Blandford1978}. Besides the non-thermal radiation from the jet, which typically dominates the radiative output of the source, there are several other sources of radiation in the blazar environment, e.g., the accretion disk, the broad line region (BLR), and the dusty torus \citep[for the BLR emission of \txs, see][]{Padovani:2019xcv}. 

\begin{figure}
\includegraphics[width=\linewidth]{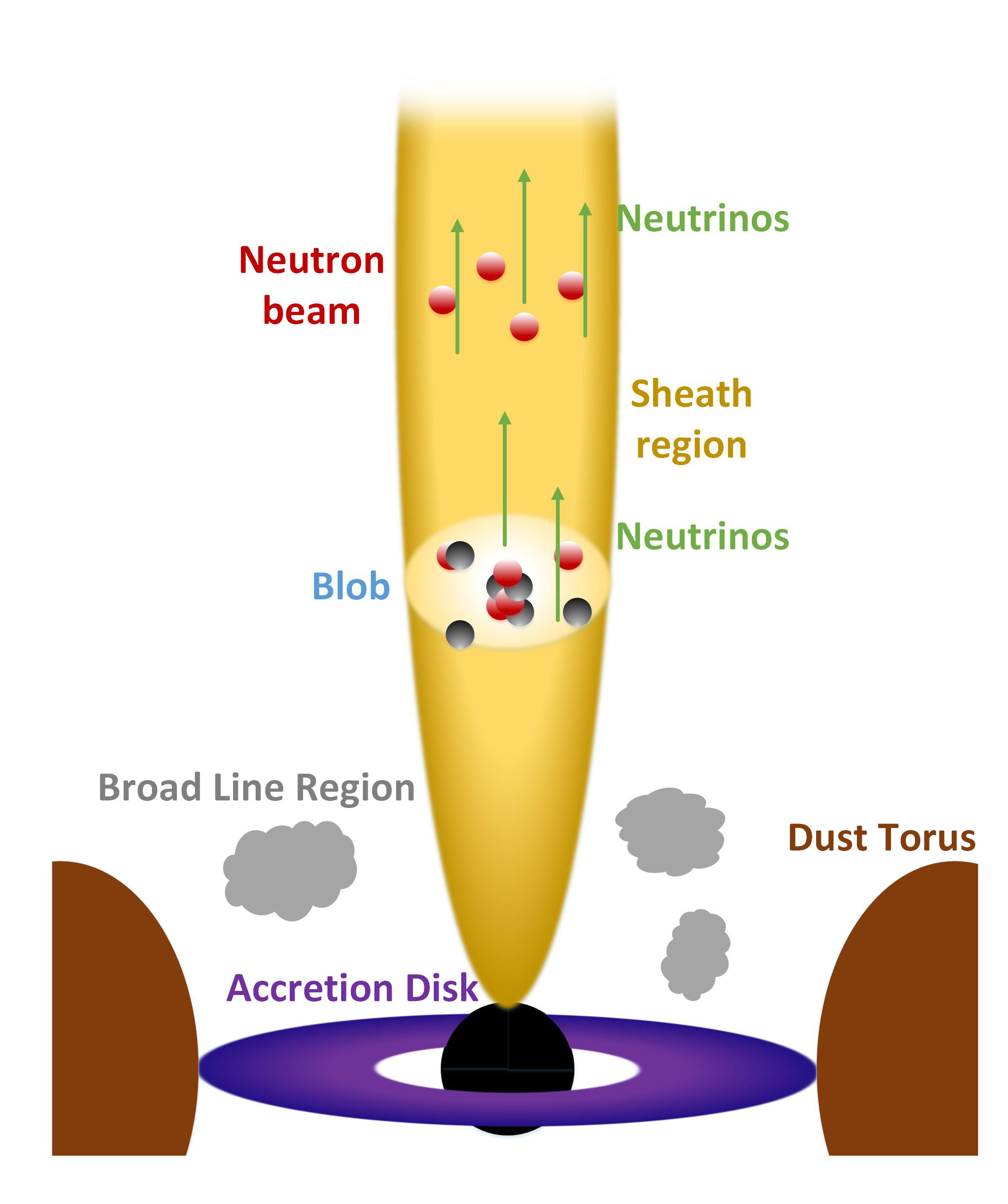}
\caption{A schematic view of the neutral beam model for blazars (not in scale). Protons and heavier nuclei are accelerated in a localized region of the blazar jet (blob), where they interact with various photon fields in the blazar environment to produce high-energy photons, electron-positron pairs, neutrinos, and neutrons. While pairs may radiate away their energy and high-energy photons may be attenuated before they escape the blob, neutrinos and neutrons can freely escape. Neutrons continue interacting with external photons on larger scales, as they propagate along the jet forming a collimated beam ($\theta\sim 1/\Gamma$), to produce more high-energy neutrinos and pairs. The neutrino emission is still beamed, but the associated cascade emission from the pairs will be diminished because of angular and time spreads.}
\label{fig:Blazar}
\end{figure}

Variable broadband blazar emission is believed to originate in the jet, but the location and dissipation mechanisms remain unclear \citep[for a recent review, see ][]{Bindu2019}. Here, we model the region, wherein accelerated particles are injected, as a spherical blob of radius $R^\prime_b$ that contains a tangled magnetic field of strength $B^\prime$ and moves with a bulk Lorentz factor $\Gamma$ (see Figure~\ref{fig:Blazar} for an illustration). Henceforth, primed quantities are measured in the rest frame of the blob (or jet). The co-moving spectral luminosity of all emission produced in the blob appears boosted in the observer's frame, i.e., $\varepsilon_i L_{\varepsilon_i}=\delta^4 \varepsilon_i^\prime L^\prime_{\varepsilon_i^\prime}$, where $\delta\equiv \Gamma^{-1}(1-\beta\cos\theta)^{-1} \approx \Gamma$; the approximation is valid for $\Gamma\gg1$ and small angles between the observer's line of sight and the jet axis  (i.e., $\theta\sim 1/\Gamma$). We also derive the isotropic-equivalent luminosity of the beam-induced neutrinos in the observer's frame via the transformation $\varepsilon_\nu L_{\varepsilon_\nu} = \delta^4\varepsilon_\nu^\prime L_{\varepsilon_\nu^\prime}^\prime$, since the produced neutrinos are assumed to be collimated within an angle $\sim 1/\Gamma$.

We assume that charged particles, including (primary) electrons, protons, and heavier nuclei are accelerated to high energies before they are injected into the blob, where they subsequently lose energy through various radiative processes. Several mechanisms of particle acceleration have been discussed in application to AGN jets, e.g., Fermi type I~\citep{Dermer:2010iz, Inoue:2016fwn}, Fermi type II~\citep{Boettcher:1998hp, Schlickeiser:2000qg,Katarzynski:2006zc}, magnetic reconnection \citep{Lovelace:1997,Giannios:2009, Petropoulou:2016, Nalewajko:2018, Christie:2019}, shear acceleration~\citep{Rieger:2004, Rieger:2007, Kimura:2017ubz}. 
In all scenarios, the acceleration efficiency depends on local plasma conditions \citep[for relativistic shocks in magnetized jets, see][]{Sironi:2015, SPG:2015}.
Despite the specifics of the acceleration process, the resulting particle energy spectrum can be phenomenologically  described by a power law terminated by a high-energy exponential cutoff, which for a nucleus of mass number $A$ can be written as:
\begin{equation}
\varepsilon^\prime_A L_{\varepsilon^\prime_A} \propto f_A \left(\frac{\varepsilon_A^\prime}{Z}\right)^{1 - s_{\rm acc}} {\rm exp} \left(-\frac{\varepsilon_A^\prime}{ \varepsilon_{A, \rm max}^\prime}\right),
\end{equation}
where the normalization is determined by the total cosmic-ray injection luminosity $L_{\rm CR}^\prime$, $s_{\rm acc}$ is the power-law index, $f_A$ is the number fraction of accelerated nuclei, $Z$ is the charge number, $\varepsilon_{p, \max}^\prime$ is the maximum proton energy, and $\varepsilon^\prime_{A, \max} = Z \varepsilon^\prime_{p, \rm max}$.  
In the following, we consider hard power-law energy spectra for the accelerated nuclei with $s_{\rm acc} = 1$ and adopt $\varepsilon^\prime_{p,\min}=1$~GeV; similar results are obtained for any other choice of the power law index as long as $s_{\rm acc}<2$.

Radiation fields that are external to the jet can affect the photo-hadronic interaction rates of protons and nuclei~\citep{Dermer:2000gu, Murase:2014foa, Dermer:2014vaa, Petropoulou:2015swa,Reimer:2018vvw,Padovani:2019xcv}. In this study, we consider an arbitrary external isotropic radiation field as the main target for photo-hadronic interactions of nuclei/protons in the blob and neutrons in the beam. We assume that the differential photon number density has a power-law-like spectrum\footnote{For the purposes of the analytical calculations presented in Section~\ref{sec:analytics}, we simply consider a power-law target photon field.}: 
\begin{equation}
\varepsilon_\gamma^\prime n_{\varepsilon_{\gamma}^\prime} = n_{\rm ex, 0}^\prime\left(\frac{\varepsilon_\gamma^\prime}{\varepsilon_{\gamma, \max}^\prime} \right)^{1 -s_{\rm ex}} \!\!\!\!\!\!\!\!{\rm exp} \left(-\frac{\varepsilon_\gamma^\prime}{\varepsilon_{\gamma, \max}^\prime} \right) {\rm exp} \left(-\frac{\varepsilon_{\gamma, \rm min}^\prime}{\varepsilon_{\gamma}^\prime}\right),
\label{eq:nex}
\end{equation}
where $n_{\rm ex, 0}^\prime$ is a normalization factor determined by the co-moving energy density $u^\prime_{\rm ex}$, $s_{\rm ex}$ is the photon index, and $\varepsilon_{\gamma, \rm min}^\prime$, $\varepsilon_{\gamma, \rm max}^\prime$ are the minimum and maximum photon energies, respectively. We further assume that the co-moving energy density of the external photon field is much larger than any of the non-thermal photon emission produced within the blob, so that we can safely neglect the latter in our neutrino calculations. We discuss possible origins of the external radiation in Section~\ref{sec:5}.

\subsection{Neutral beam and neutrino production}\label{sec:analytics}
Here we discuss the physics of neutrino production with analytical expressions, although the calculations are performed numerically. 

Neutrinos are  natural by-products of the photomeson production process of nuclei inside the blob:
\begin{equation}
\varepsilon_\nu^\prime L_{\varepsilon_\nu^\prime}^\prime |_{\rm blob} \approx \frac{3}{8} \sum_A {f}^{\rm (mes)}_{A\gamma} \varepsilon_A^\prime L_{\varepsilon_A^\prime}^\prime,
\end{equation}
where ${f}^{\rm (mes)}_{A\gamma} (\varepsilon_A^\prime) \sim f_{p\gamma} (\varepsilon_A^\prime / A)$ (that is assumed to be less than unity) is the energy loss efficiency of the photomeson process for nuclei~\citep{Murase:2010gj, Zhang:2018agl} and $L_{\varepsilon_A^\prime}^\prime \equiv \varepsilon_A^\prime dN^\prime / d \varepsilon_A^\prime$. The proton photomeson energy loss efficiency can be written as \cite[e.g.,][]{Murase:2018iyl}:
\begin{eqnarray}
f_{p\gamma}(\varepsilon_p^\prime) &\approx& \frac{2 \hat{\sigma}_{p\gamma}}{1+s_{\rm ex}} R_b^\prime n_{\rm ex, 0}^\prime \left(\frac{\varepsilon_p^\prime}{\varepsilon_{p, 0}^\prime}\right)^{s_{\rm ex}-1} \nonumber \\ &\sim& 3 \times 10^{-3} \eta_{p\gamma}[s_{\rm ex}] R_{b, 15}^\prime (\varepsilon_p^\prime / \varepsilon_{p, 0}^\prime)^{s_{\rm ex} - 1},
\label{eq:fpg}
\end{eqnarray}
where equation (\ref{eq:nex}) is used, $\eta_{p\gamma}[s_{\rm ex}] = 2 / (1 + s_{\rm ex})$, $\hat{\sigma}_{p\gamma} \simeq 0.7 \times 10^{-28} \rm~cm^2$ is the effective cross section for photomeson interactions of protons, $\varepsilon_{p, 0}^\prime = 0.5 m_p c^2 \bar{\varepsilon}_\Delta / \varepsilon_{\gamma, \rm max}^\prime$, $\bar{\varepsilon}_\Delta \sim 0.34\rm~GeV$, $\varepsilon_{\gamma, \rm min}^\prime = 10^{-2}\rm~eV$, $\varepsilon_{\gamma, \rm max}^\prime = 10^2\rm~eV$, and $n_{\rm ex, 0}^\prime \sim 3.5 \times 10^8\rm~cm^{-3}$ corresponding to $u_{\rm ex}^\prime = 10\rm~erg~cm^{-3}$.

Photomeson production by nucleons or nuclei also leads to neutrons as well as neutrinos. The beam of neutrons and secondary particles is collimated with an opening angle of $\sim 1/\Gamma$, where $\Gamma \sim \delta$. The co-moving luminosity of the escaping neutral beam can be estimated as follows~\citep{Murase:2018iyl}: 
\begin{equation}
\varepsilon_n^\prime L_{\varepsilon_n^\prime}^\prime \approx \sum_A \zeta_{n} f_{A\gamma} \varepsilon_A^\prime L_{\varepsilon_A^\prime}^\prime,
\end{equation}
where for simplicity we consider only the photodisintegration process as a source of neutrons\footnote{In the numerical calculations we also consider neutron production through the photomeson production process.}, and $\zeta_{n}\sim1/2$ is the fraction of neutrons in the emitted nucleons, to be determined by numerical simulations.
The effective optical depth for the photodisintegration of nuclei $f_{A\gamma}^\prime$ can be estimated as ~\citep{Murase:2010gj, Zhang:2017hom}:
\begin{eqnarray}
f_{A\gamma}(\varepsilon_A^\prime) &\approx& \frac{2 \hat{\sigma}_{A\gamma}}{1+s_{\rm ex}} R_b^\prime n_{\rm ex, 0}^\prime \left(\frac{\varepsilon_A^\prime}{\varepsilon_{A, 0}^\prime}\right)^{s_{\rm ex}-1} \nonumber \\ &\sim& 7 \times 10^{-2} \eta_{A\gamma}[s_{\rm ex}] R_{b, 15}^\prime (\varepsilon_A^\prime / \varepsilon_{A, 0}^\prime)^{s_{\rm ex} - 1},
\end{eqnarray}
where equation (\ref{eq:nex}) is used, $\eta_{A\gamma}[s_{\rm ex}]=2/(1+s_{\rm ex})$, $\hat{\sigma}_{A\gamma} \equiv \sigma_{\rm GDR} \Delta \bar{\varepsilon}_{\rm GDR} / \bar{\varepsilon}_{\rm GDR} \simeq 1.7 \times 10^{-27}\rm~cm^2$ is the effective photodisintegration cross section for helium nuclei, $\varepsilon_{A, 0}^\prime = 0.5 m_A c^2 \bar{\varepsilon}_{\rm GDR} / \varepsilon_{\gamma, \rm max}^\prime$, $\bar{\varepsilon}_{\rm GDR} = 0.925 A^{2.433}$ for $A \leq 4$~\citep{Murase:2010gj}, and $\varepsilon_{\gamma, \rm max}^\prime = 10^2 \rm~eV$ is used to get the numerical value in the above estimate.

In addition, neutrinos can be produced from the interaction of the escaping neutrons with external radiation fields, with an estimated luminosity of:
\begin{equation}
\varepsilon_\nu^\prime L_{\varepsilon_\nu^\prime}^\prime |_{\rm beam} \approx \frac{3}{8} f_{n\gamma} \varepsilon_n^\prime L_{\varepsilon_n^\prime}^\prime,
\end{equation}
where $f_{n\gamma}$ is the energy loss efficiency of the escaping neutron beam, which is assumed to be less than unity. 
The efficiency can be written as:
\begin{eqnarray}
f_{n\gamma}(\varepsilon_n^\prime) &\approx& \frac{2 \hat{\sigma}_{n\gamma}}{1+s_{\rm ex}} R_{\rm ext}\frac{n_{\rm ex, 0}^\prime}{\Gamma}\left(\frac{\varepsilon_n^\prime}{\varepsilon_{n, 0}^\prime}\right)^{s_{\rm ex}-1} \nonumber \\ &\sim& 1 \times 10^{-1} \eta_{n\gamma}[s_{\rm ex}] R_{\rm ex, 17.6} \Gamma_{1}^{-1} (\varepsilon_n^\prime / \varepsilon_{n, 0}^\prime)^{s_{\rm ex} - 1},
\end{eqnarray}
where $R_{\rm ex}$ is the size of the external radiation field (as measured in the AGN rest frame), $\eta_{n\gamma}[s_{\rm ex}]=2/(1 + s_{\rm ex})$, $\hat{\sigma}_{n\gamma}\sim \hat{\sigma}_{p\gamma}$ is the effective cross section for photomeson interactions of neutrons, $\varepsilon_{n, 0}^\prime=0.5m_nc^2 \bar{\varepsilon}_\Delta / \varepsilon_{\gamma, \rm max}^\prime$, and $\varepsilon_{\gamma, \rm max}^\prime = 10^2\rm~eV$ is used in the final estimate.

The luminosity ratio of beam-induced neutrinos and  blob-induced neutrinos (for a single species of nuclei) can be then estimated as:
\begin{eqnarray}
\xi_\nu &\equiv&  \frac{\varepsilon_\nu^\prime L_{\varepsilon_\nu^\prime}^\prime |_{\rm beam}}{ \varepsilon_\nu^\prime L_{\varepsilon_\nu^\prime}^\prime |_{\rm blob}} \nonumber 
\approx \frac{f_{n\gamma}}{ {f}^{\rm (mes)}_{A\gamma}} \frac{\varepsilon_n^\prime L_{\varepsilon_n^\prime}^\prime}{\varepsilon_A^\prime L_{\varepsilon_A^\prime}^\prime} \nonumber \\ 
&\approx& \frac{f_{n\gamma}}{{f}^{\rm (mes)}_{A\gamma}} \zeta_{n} f_{A\gamma}^\prime 
\sim  \frac{f_{n\gamma}[\varepsilon_n^\prime]}{f_{p\gamma} [\varepsilon_A^\prime / A]} \zeta_{n} f_{A\gamma} [\varepsilon_A^\prime].
\label{eq:xi}
\end{eqnarray}
For the same target field, the ratio $f_{n\gamma}[\varepsilon_n^\prime]/f_{p\gamma} [\varepsilon_A^\prime / A]$ depends on the relative size of the blob and the external radiation field as $\propto R_{\rm ex} / \Gamma R_b^\prime$, with its exact value determined by the details of the photodisintegration processes inside the blob. For example, if $R_{\rm ex} / \Gamma R_b^\prime \sim 40$, then we have $\xi_\nu \sim 3 \, R_{\rm ex, 17.6} \Gamma_1^{-1} R_{b, 15}^\prime (\zeta_n / 1) (f_{A\gamma} [\varepsilon_A^\prime] / 0.07)$. 
Thus, for certain parameters the neutrino emission produced from the neutral beam can be several times larger than blob's neutrino emission~\citep{Murase:2018iyl}.

\subsection{Constraints from electromagnetic cascades}
Along with neutrinos, the  photomeson production process leads to the generation of relativistic electron-positron pairs (henceforth, we will refer to them simply as pairs) via the decay of charged pions and of $\gamma$-ray photons from the decay of neutral pions. Another source of relativistic pairs is the BH process of protons and nuclei \citep[for its importance in blazars, see, e.g.,][]{Petropoulou:2014rla,Murase:2014foa,Petropoulou2015,Keivani:2018rnh,Murase:2018iyl}.
The effective optical depth to the BH pair-production process can be written as~\citep{Murase:2018iyl}:
\begin{equation}\label{eq:BH}
f_{{\rm BH}, A}\left(\varepsilon_A^\prime\right)\approx\frac{2\hat{\sigma}_{{\rm BH}, A}}{1 + s_{\rm ex}} R_b^\prime n_{\rm ex, 0}^\prime \left(\frac{\varepsilon_A^\prime}{\varepsilon_{\rm BH, 0}}\right)^{s_{\rm ex} - 1},
\end{equation}
where equation (\ref{eq:nex}) is used, $\hat{\sigma}_{{\rm BH}, A}\approx(Z^2/A) \hat{\sigma}_{{\rm BH, p}} \simeq 8\times 10^{-31} (Z^2/A) \rm~cm^{2}$ is the photopair cross section of  nuclei, and $\tilde{\varepsilon}_{\rm BH,0}^\prime=\bar{\varepsilon}_{\rm BH}m_pc^2/2\varepsilon'_{\gamma,\rm max}$ with $\bar{\varepsilon}_{\rm BH} \sim 10(2m_ec^2)\sim 10\rm~MeV$ \citep{1992ApJ...400..181C, Murase:2018iyl}. 

The secondary pairs produced within the blob can radiate away their energy via synchrotron and inverse Compton scattering (ICS) processes before they escape from the emitting region. Similarly, $\gamma$-ray photons from neutral pion decays can be attenuated by soft radiation in the blob, resulting in the production of more relativistic pairs. The net result is the development of an electromagnetic cascade within the blob, whose flux is limited by available multi-wavelength observations \citep[see, e.g.,][]{Keivani:2018rnh, Gao:2019NatAs}.

We argue that the emission from secondaries produced by the neutral beam can be suppressed for two reasons. First, neutrons do not pair-produce via the BH process, which turns out to be the most important source for relativistic pairs, as we show later in Section~\ref{sec:4}. Second, the emission of pairs, including both BH pairs and those produced from $\gamma$-ray attenuation, should largely be isotropized in the larger scale jet where the magnetic field is weaker \citep[for details, see][]{Murase:2018iyl}. 
Because of the resulting anisotropic cascades, the radiation produced by the secondaries (as seen in the observer's frame) can readily be suppressed via both angular spreading and time delay. 
The delayed synchrotron cascade emission originating from pairs with $\sim{10}^{3}$~GeV is unlikely to be observed, although $\gamma$-ray emission from ultrahigh-energy pairs could in principle be detected as synchrotron ``pair-echo'' emission \citep{Murase:2011yw,Dermer:2012rg}.  

\section{Numerical approach}\label{sec:3}
\begin{figure}
\includegraphics[width=0.5\textwidth]{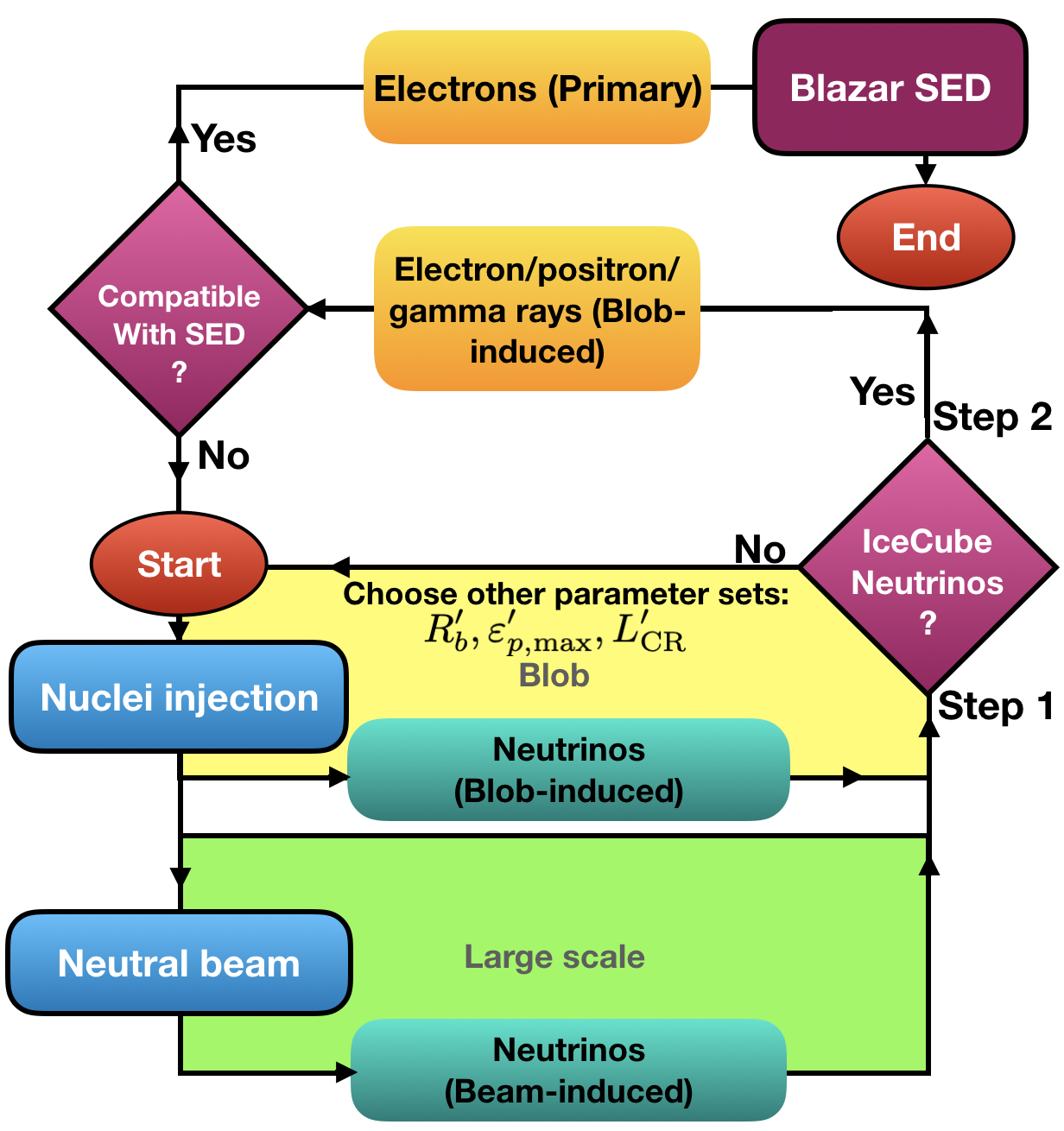}
\caption{Flowchart demonstrating the two-step numerical approach of our study. We begin by injecting nuclei into the blob. We then calculate the neutrino emission from the blob and the beam and compare to the 2014--2015 neutrino flux measured by IceCube. If the integrated model-predicted neutrino flux is not consistent with the observed one (Step 1), we choose another parameter set. Otherwise, we check if the accompanying cascade emission in the blob overshoots the electromagnetic data (Step 2). If it does, then we choose another input parameter set and repeat Step 1. Otherwise, we check if the model can explain the blazar SED by taking into account the emission of primary electrons in the blob.}
\label{fig:flowchart}
\end{figure}

In this section, we describe our methods for the computation of multi-messenger emission from the neutral beam and the blob. A flowchart describing our numerical approach is shown in Figure~\ref{fig:flowchart}.

We first consider photo-hadronic interactions of nucleons and nuclei in the blob and of the neutral beam in the jet, and derive energy spectra of neutrinos, pairs, and $\gamma$-rays from pion decays. 
We utilize the publicly available Monte--Carlo code 
{\sc CRPropa-3.0}~\citep{Batista:2016yrx} 
which takes into account photodisintegration interactions~\citep{Rachen1996},  photomeson production~\citep{Mucke:1999yb}, and BH pair-production~\citep{1992ApJ...400..181C} processes of protons and nuclei\footnote{Synchrotron cooling of secondary pions and muons, which could suppress the neutrino flux, becomes important only for $B^\prime \gtrsim 10^4$~G \citep[e.g.,][]{Murase:2005hy, Baerwald2011, PDG2014, Kimura:2017kan, Zhang:2018agl}, which is unlikely to realize in sub-pc scale blazar jets.}.
All calculations are performed in the rest frame of the blob (or jet). We solve the rectilinear propagation of protons and nuclei for a travel distance of $R'_b$~\citep[see][for the application to engine-driven supernovae]{Zhang:2018agl}. We assume that all charged particles (protons and nuclei) cannot escape due to magnetic confinement in the blob and lose energy via adiabatic losses~\citep[e.g.,][]{Dermer:2012rg}. 
On the other hand, neutrons are free from magnetic confinement, and the escaping relativistic neutrons will continue to interact with the external photon field (until they reach the edge of the extension of the exteral photon field) and produce more neutrinos. A similar approach was adopted by \cite{Dermer:2012rg}. At the end of the calculation, we compute the neutrino flux from the blob and the beam and compare it with the IceCube data.

In the second step, we compute electromagnetic cascade emission induced by secondaries injected in the blob as well as the emission of primary electrons with the goal of explaining the observed SED.
We use the time-dependent numerical code described in \cite{Dimitrakoudis2012} to compute the emission of the cascade in the blob. 
In particular, we solve the kinetic equations describing the evolution of photons and pairs, using as source terms in the relevant equations the production rates of secondary pairs and photons as derived from {\sc CRPropa-3.0}. We take into account all the relevant loss terms for both particle species (i.e., synchrotron radiation, ICS, and $\gamma \gamma$ absorption). Although the adopted numerical approach treats the feedback between photons and pairs produced in the blob self-consistently \citep{Dimitrakoudis2012, Petropoulou2014}, it is not designed to treat the feedback of the cascade photons on the photomeson production and photodisintegration rates. However, the latter effects are negligible as long as the external photon density is much higher than the density of locally produced photons. 

Using the injection rates and external photon field properties from the first step, we  compute the cascade emission  for indicative values of the blob magnetic field strength. If the cascade emission
overshoots the electromagnetic data, then we discard the model. If, however, the cascade emission is consistent with these data for some $B^\prime$ value, we then consider the emission of primary electrons accelerated in the blob along with the protons and nuclei.  We finally check if the resulting photon spectrum can explain the blazar SED.  

\subsection{Model parameters}
The plasma composition in AGN jets cannot be probed directly and, as a result, remains largely unknown. In addition to this, the acceleration efficiency of different particle species, which will ultimately determine the composition of particles injected into the blob, is expected to depend on the plasma conditions as well as on the acceleration process itself. Detailed numerical studies on the acceleration efficiency of different nuclei are sparse. 
For example, \citet{Caprioli:2017oun} demonstrated that the non-thermal tail of nuclei is enhanced by ${(A /Z)}^2$, for the efficient diffusive acceleration at non-relativistic shocks and singly-ionized material. Albeit informative, at present, such studies are not conclusive and cannot be directly applied to relativistic magnetised outflows of blazars.
We therefore treat the composition of accelerated nuclei that are injected into the blob as a free parameter. In particular, we consider a scenario where the injected nuclei are composed of protons and helium, namely two of the most abundant elements in the Universe. The accelerated helium-to-proton number ratio ($f_{\rm He}/f_{\rm P}$) is a free model parameter to be constrained by the combined neutrino and electromagnetic data. We use $f_{\rm He}/f_{\rm P}=5\, \sim 4^2\times(0.24/0.76)$ as our benchmark value, where 0.24 and 0.76 are solar mass fractions of helium and protons.  
\begin{table}
\caption{\label{tab:tab1} A list of input model parameters.}
\begin{ruledtabular}
\begin{tabular}{llll}
\textrm{Physical parameters} & \multicolumn{3}{c}{\textrm{Value}}\\
\colrule
\rule{0pt}{3ex} \\
\multicolumn{4}{l}{\textbf{Default parameters used in this work}}  \\
External photon field radius ($R_{\rm ex}$ [cm]) & \multicolumn{3}{c}{$2 \times 10^{17}$} \\
External energy density ($u_{\rm ex}^\prime$ [erg/cm$^{3}$]) & \multicolumn{3}{c}{$100$} \\
External photon spectral index ($s_{\rm ex}$) & \multicolumn{3}{c}{1.5} \\
Minimum photon energy ($\varepsilon_{\rm min}^\prime$ [eV]) & \multicolumn{3}{c}{$10^{-2}$} \\
Maximum photon energy ($\varepsilon_{\rm max}^\prime$ [eV]) & \multicolumn{3}{c}{$2 \times 10^2$} \\
Blob Lorentz factor ($\Gamma$) & \multicolumn{3}{c}{$10$} \\
Minimum proton energy ($\varepsilon_{p, \rm max}^\prime$ [GeV]) & \multicolumn{3}{c}{$1$}\\
Nuclei acceleration spectral index ($s_{\rm acc}$) & \multicolumn{3}{c}{1} \\
Number ratio of accelerated nuclei ($f_{\rm He} / f_{\rm P}$) & \multicolumn{3}{c}{5} \\
\rule{0pt}{3ex} \\
\multicolumn{4}{l}{\textbf{Optimized parameters for  neutrino flux}} \\ 
Model & A & B & C \\
\hline
Maximum proton energy ($\varepsilon_{p, \rm max}^\prime$ [PeV]) & 0.6 & 0.4 & 0.6 \\
Blob radius ($R_b^\prime$ [$10^{15}$ cm]) & 1 & 1 & 6 \\
Total CR luminosity ($L_{\rm CR}$ [$10^{49}$ erg/s]) & 9 & 25 & 7 \\
\rule{0pt}{3ex} \\
\multicolumn{4}{l}{\textbf{Additional parameters for SED}} \\
Model & A & B & C \\
\hline
Magnetic field strength ($B^\prime$ [G]) & 80 & 80 & 80 \\
Minimum electron energy ($\varepsilon_{e, \rm min}^\prime$ [GeV]) & 0.3 & 0.3 & 0.2 \\
Maximum electron energy ($\varepsilon_{e, \rm max}^\prime$ [GeV]) & 0.5 & 0.5 & 0.3  \\
Spectral index ($s_{\rm acc}$) & 1.9 & 1.9 & 1.9  \\
Total electron luminosity ($L_{\rm e}$ [$10^{46}$ erg/s]) & 5.8 & 5.8 & 5.8 \\
\end{tabular}
\end{ruledtabular}
\tablecomments{The models A-C are introduced in Section~\ref{sec:4-1}.}
\end{table}

We also treat the external photon field as a free parameter that optimizes the beam/blob neutrino production. The external radiation is constrained by the requirement that its flux (in the observer's frame) is out-shined by the non-thermal emission of the blob, namely $L_{\rm ex} \approx 4\pi R_{\rm ex}^2 c u_{\rm ex}^\prime / \Gamma^2 \lesssim 10^{46}\rm~erg~s^{-1}$.
Note that this is a necessary requirement for any type of additional radiation fields (from e.g., the sheath region of the jet). 
As we discussed in Section~\ref{sec:analytics}, the luminosity ratio of beam-induced neutrinos and blob-induced neutrinos is $\xi_\nu \propto R_{\rm ex} u_{\rm ex}^\prime \propto \Gamma^2 L_{\rm ex} / R_{\rm ex} $. For fixed $L_{\rm ex}$ and $\Gamma$, the ratio $\xi_\nu$ increases for smaller values of $R_{\rm ex}$ and higher values of $u_{\rm ex}^\prime$ (i.e., more compact photon regions). 
Assuming the observed typical neutrino energy is $\varepsilon_\nu = 1\rm~PeV$, then the corresponding target photon energy measured in the blob comoving frame is $\varepsilon_\gamma^\prime \sim 10^2 (\Gamma / 10) \rm~eV$. Based on these considerations, we choose the following parameters values for our default external photon field model: $u_{\rm ex}^\prime = 100\rm~erg~cm^{-3}$, spectral index $s_{\rm ex} = 1.5$, maximum photon energy $\varepsilon_{\rm max}^\prime = 2 \times 10^2\rm~eV$, minimum photon energy $\varepsilon_{\rm min}^\prime = 10^{-2}\rm~eV$, and  radius $R_{\rm ex} = 2 \times 10^{17}\rm~cm$ (see Table~\ref{tab:tab1}). 

The other free parameters that are required to explain the neutrino data are: the maximum proton energy $\varepsilon_{p, \rm max}^\prime$, the power-law index of accelerated nuclei $s_{\rm acc}$, the total CR luminosity $L_{\rm CR}^\prime$, the blob radius $R_b^\prime$, and the blob Lorentz factor $\Gamma$. 
In order to compute electromagnetic emissions from the blob, we also need to know the co-moving magnetic field strength $B^\prime$ and the properties of primary electrons. 
Although our model is composed of two regions for neutrino production in the blazar jet, the number of free  parameters is considerably smaller than in typical two-zone models, because the two zones (i.e., beam and blob) are physically coupled to each other.

\section{\label{sec:4} Neutrino emission from \txs}
In this section, we present our results for the multi-messenger emission from \txs for the period 2014--2015. 
\subsection{Parameter space search}\label{sec:4-1}
As shown in Figure~\ref{fig:flowchart}, we compute high-energy neutrino emission for different combinations of $\varepsilon_{p, \rm max}^\prime$, $R_b^\prime$, and $L_{\rm CR}^\prime$. In particular, for each pair of values $(\varepsilon_{p, \rm max}^\prime, R_b^\prime)$, we determine $L_{\rm CR}^\prime$ so that the model-predicted neutrino flux at its peak energy ($\varepsilon_\nu^{\rm peak}$) is $\varepsilon_\nu^{\rm peak} F_{\varepsilon_\nu^{\rm peak}} = 5 \times 10^{-11}\rm~erg~cm^{-2}~s^{-1}$. For each set of $\varepsilon_{p, \rm max}^\prime$, $R_b^\prime$, and $L_{\rm CR}^\prime$ values, we also compute  the following  quantities: the luminosity ratio of beam-induced neutrinos to blob-induced neutrinos at the peak energy $\xi_\nu$, the observed peak neutrino energy $\varepsilon_\nu^{\rm peak}$, 
the total isotropic-equivalent luminosity of BH pairs $L_{\rm BH}$, the total iostropic-equivalent $\gamma$-ray luminosity from photomeson and photodisintegration interactions $L_\gamma$, and  the number of through-going muon neutrinos and anti-neutrinos, $\mathcal{N}_{\nu_\mu+\bar{\nu}_\mu}$. 

Using the energy-dependent effective area $A_{\rm eff}(\varepsilon_{\nu_{\mu}})$ of the IceCube point-source search analysis in the direction of \txs~\citep{IceCube:2018cha}\footnote{Available online at \url{https://icecube.wisc.edu/science/data/TXS0506-point-source}} we find:
\begin{equation}
\mathcal{N}_{\nu_{\mu}+\bar{\nu}_\mu} =\frac{1}{3} 
T_w\int_{\varepsilon_{\nu,{\rm min}}}^{\varepsilon_{\nu,{\rm max}}} {\rm d} \varepsilon_{\nu} \,  A_{\rm eff}(\varepsilon_{\nu}) F_{\varepsilon_{\nu}} 
\label{eq:Nnu}
\end{equation}
where $T_w=158$~days, $\varepsilon_{\nu,{\rm min}}$= 32~TeV and $\varepsilon_{\nu,{\rm max}}$= 3.6~PeV are respectively the minimum and maximum energies considered for the calculation (based on the lower energy limit of the IceCube analysis), and $F_{\varepsilon_{\nu}}$ is the all-flavor (differential in energy) neutrino flux of the model. We assume vacuum neutrino mixing and use $1/3$ to convert from the all-flavor to muon neutrino flux. 

\begin{figure*}
\centering 
\includegraphics[width=0.48\textwidth]{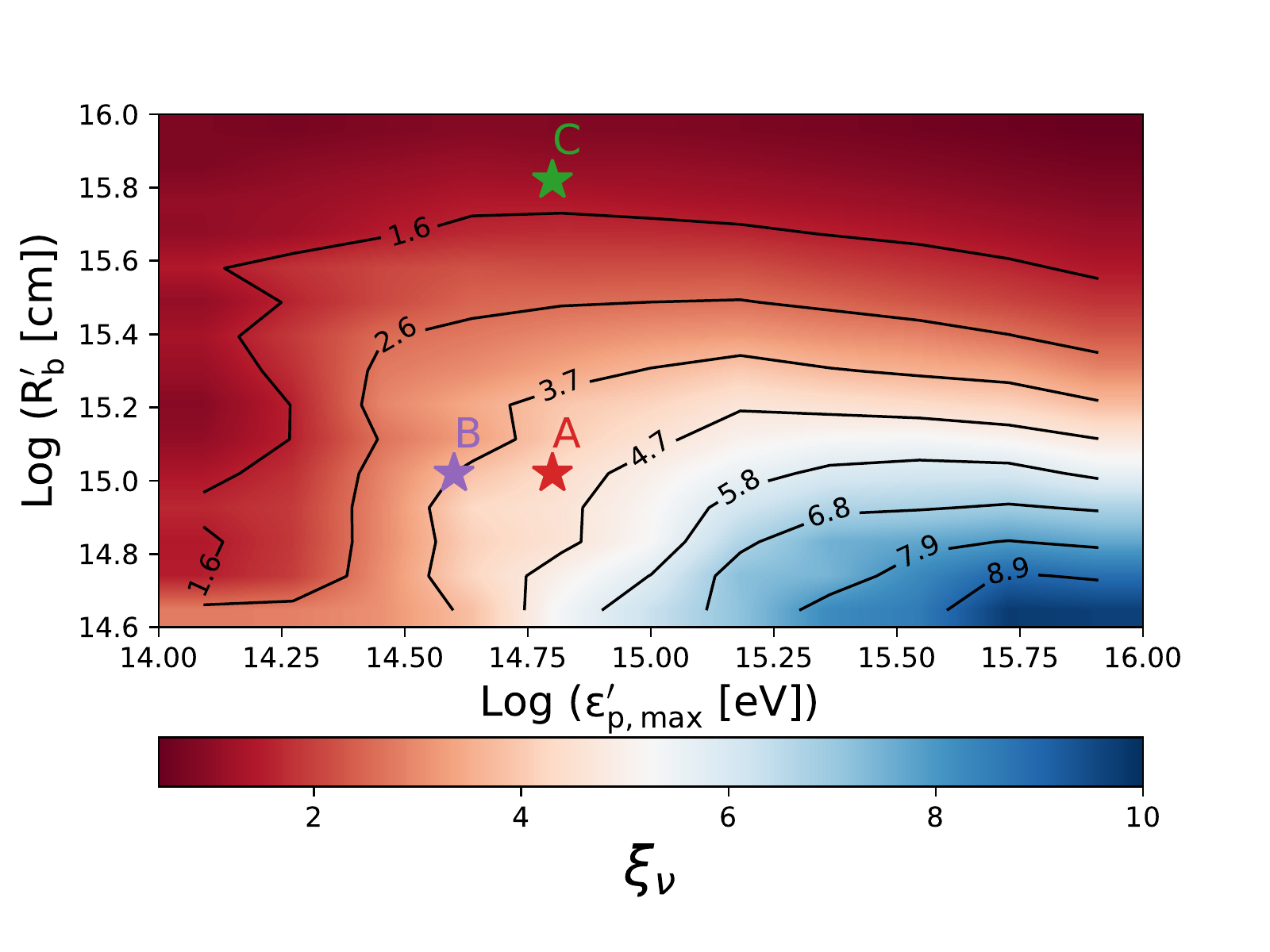}
\includegraphics[width=0.48\textwidth]{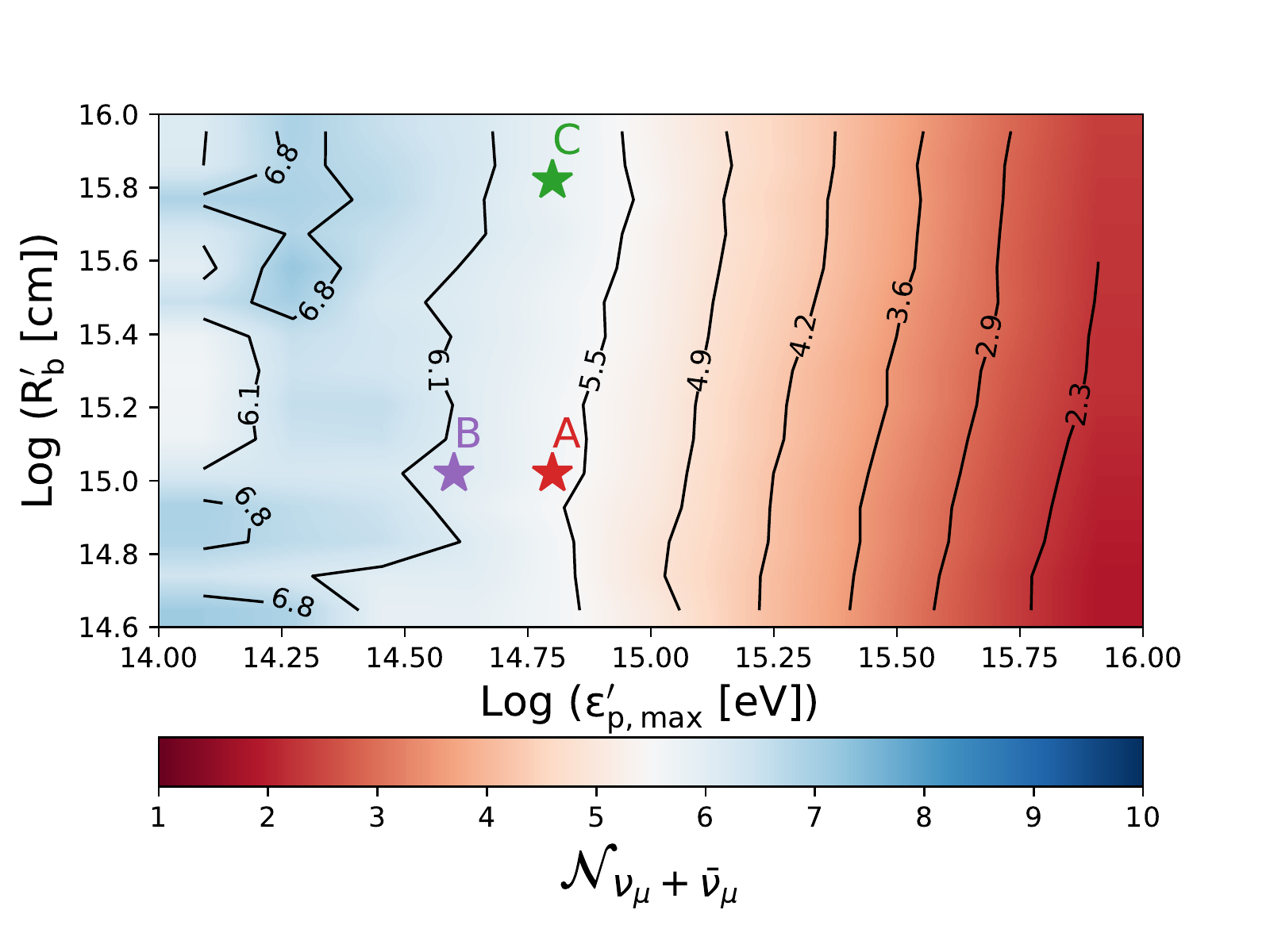}
\includegraphics[width=0.48\textwidth]{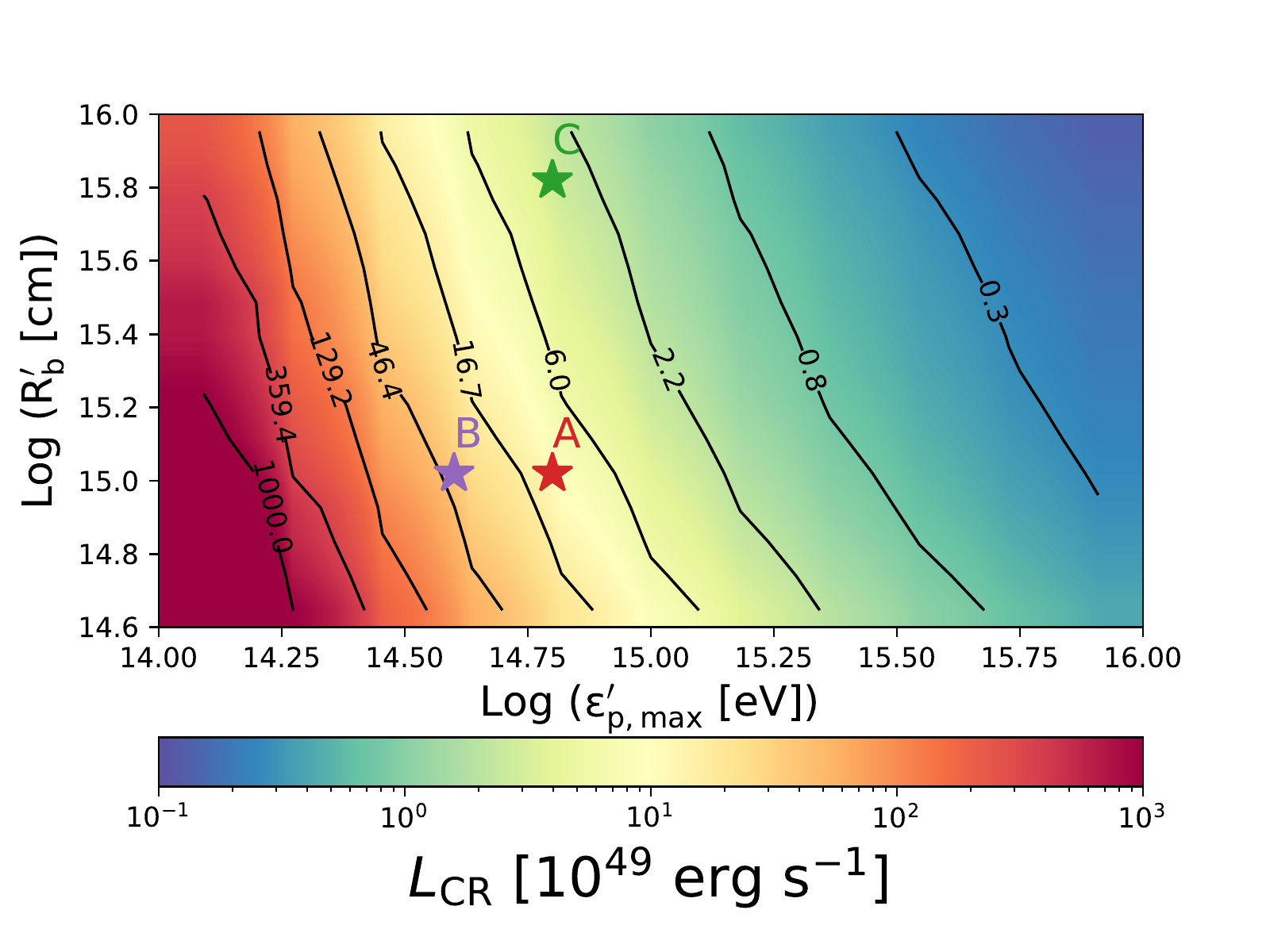}
\includegraphics[width=0.48\textwidth]{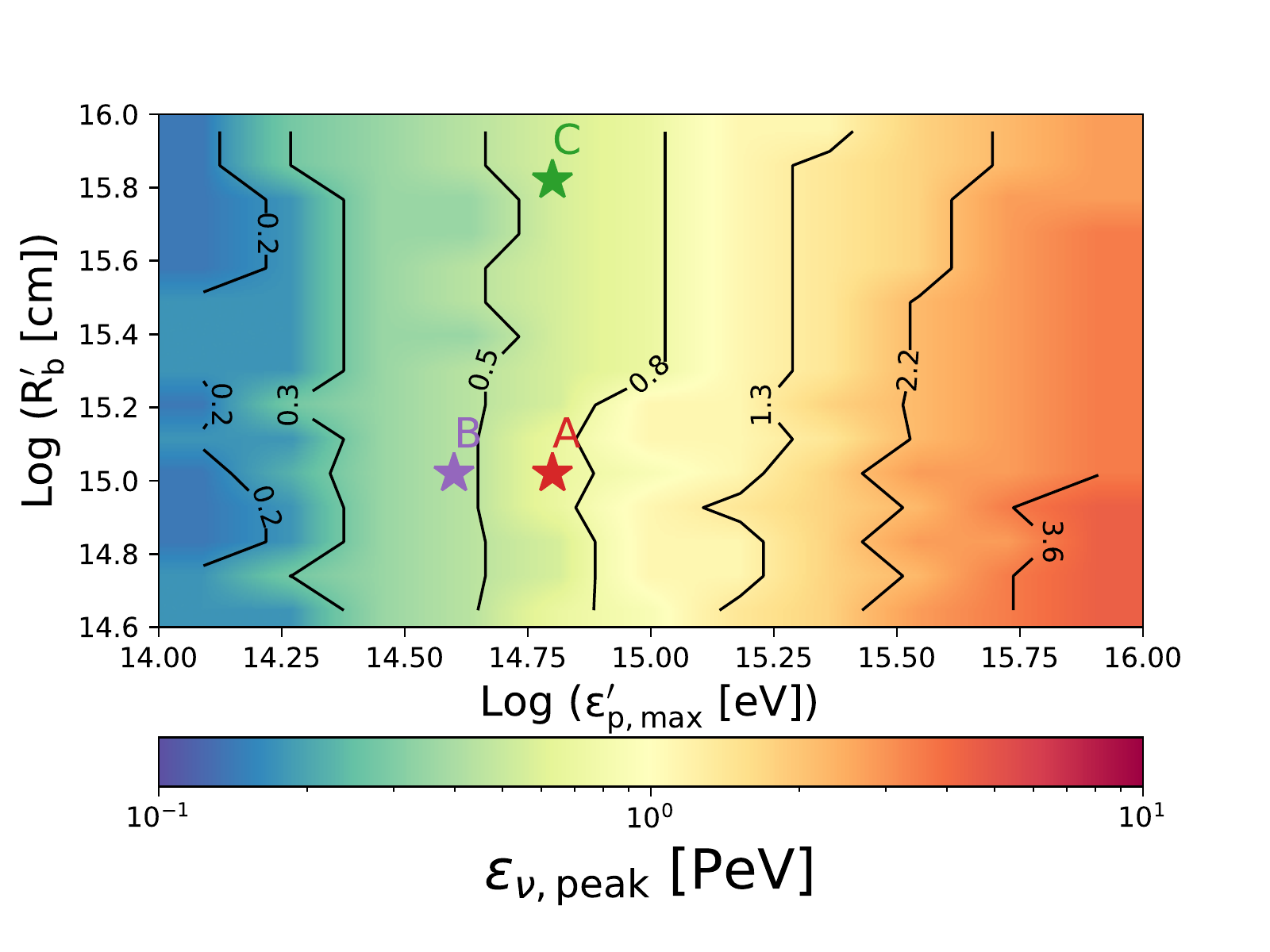}
\includegraphics[width=0.48\textwidth]{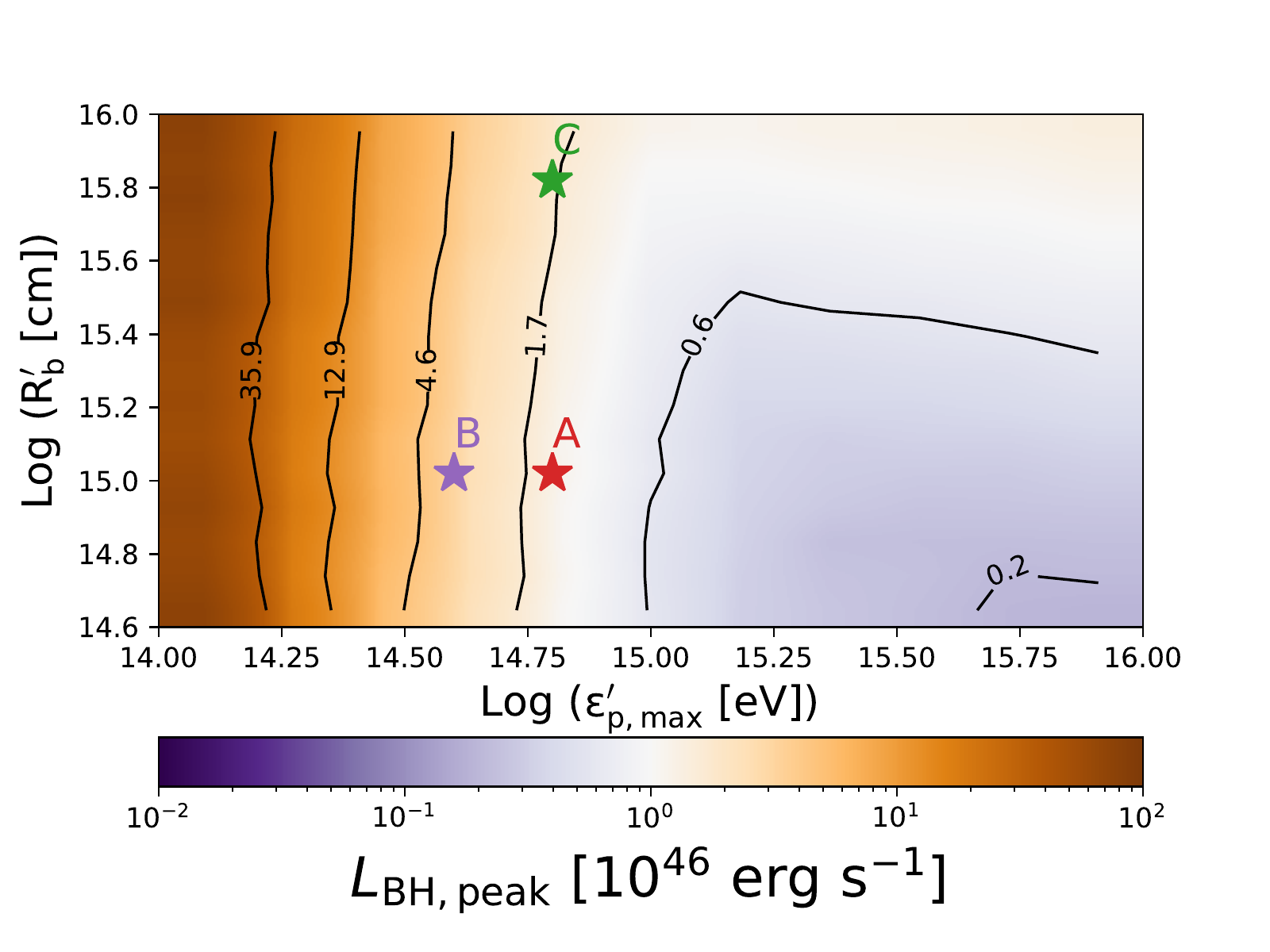}
\includegraphics[width=0.48\textwidth]{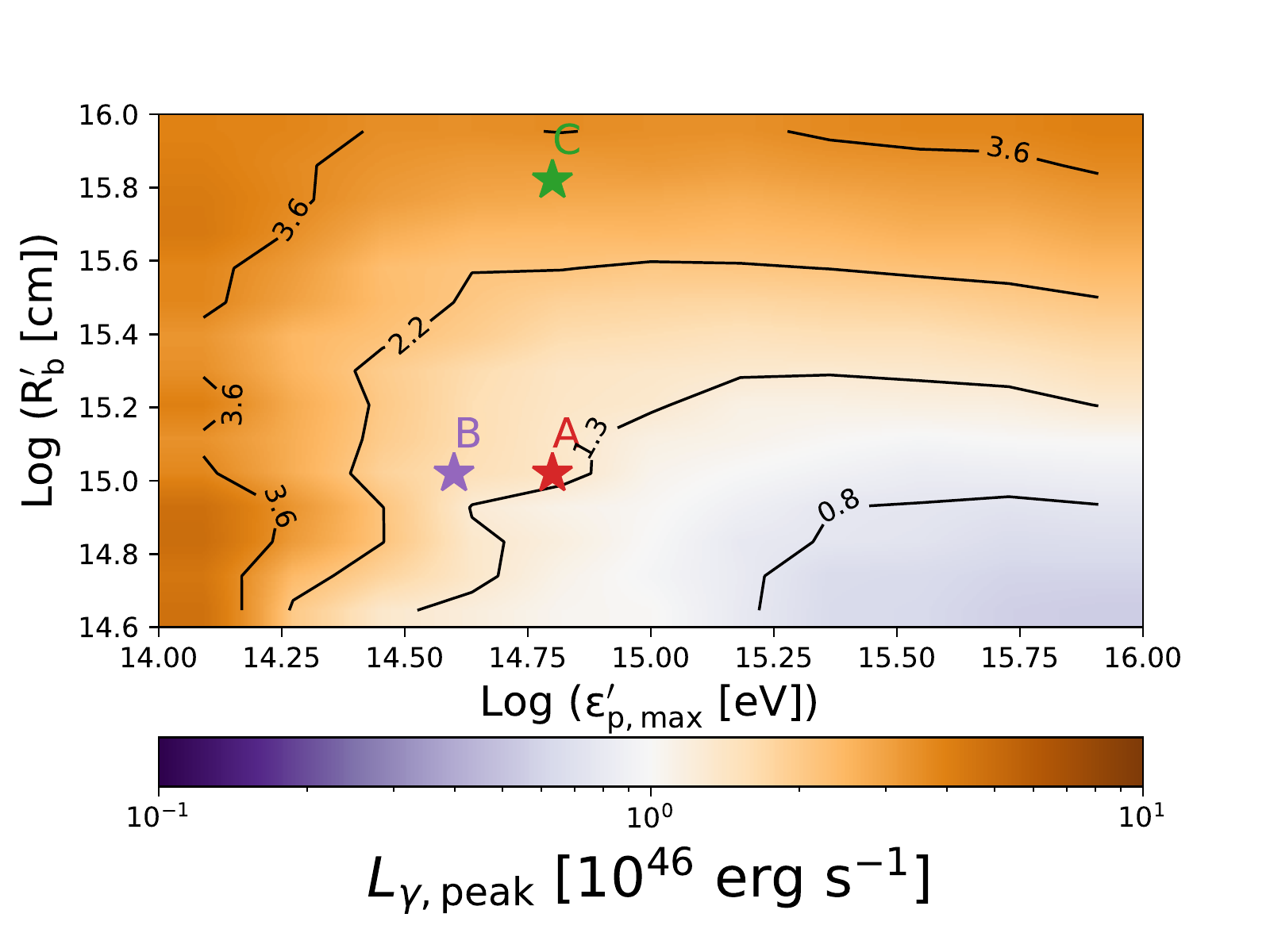}
\caption{Parameter space in the $R_b^\prime-\varepsilon_{p, \rm max}^\prime$ plane with color indicating (from top left and in clockwise order): the ratio of beam-induced and blob-induced peak neutrino luminosities $\xi_\nu$, the number of muon neutrinos and anti-neutrinos, $\mathcal{N}_{\nu_\mu + \bar{\nu}_\mu}$, the total isotropic-equivalent (observer) CR luminosity, the observed peak neutrino energy $\varepsilon_\nu^{\rm peak}$, the isotropic-equivalent (observer) peak luminosity of BH pairs, and the isotropic-equivalent (observer) peak luminosity of $\gamma$-rays. Three indicative models, which are discussed in detail in Section~\ref{sec:4-2}, are marked with stars.
\label{fig:summary_statistic_30_2d_source_ProtonHe_b}}
\end{figure*}

Our results are shown in Figure~\ref{fig:summary_statistic_30_2d_source_ProtonHe_b}.
In the top left panel, we can see that $\xi_\nu$ is sensitive to both $\varepsilon_{p, \rm max}^\prime$ and $R_b^\prime$, with higher ratios obtained for smaller $R_b^\prime$ and higher  $\varepsilon_{p, \rm max}^\prime$. These results are consistent with our analytical estimates presented in Section~\ref{sec:2}. Larger values of $\xi_\nu$ are generally preferred, because the electromagnetic cascade emission from the beam is suppressed compared to the cascade emission in the blob. Our model favors smaller blobs and more energetic nuclei. 

However, larger $\varepsilon_{p, \rm max}^\prime$ would push the observed peak neutrino energy $\varepsilon_\nu^{\rm peak}$ to higher values (see right panel in middle row) and reduce the number of neutrinos within the selected energy range (i.e., $>32$~TeV), as shown in the right panel of the top row. Given that the most probable energy of the $13\pm5$ neutrinos detected in 2014--2015 is $\sim 10-100\rm~TeV$, lower values of  $\varepsilon_{p, \rm max}^\prime$ are preferred. 

The (co-moving) maximum energy, however, cannot be much lower than $\sim 0.4$~PeV, because this would lead to much higher injection luminosities of CRs ($L_{\rm CR}$; see left panel in middle row),  secondary pairs from the BH process of nuclei ($L_{\rm BH}$; see left panel in bottom row), and photons from neutral pion decays ($L_{\gamma}$; see right panel in bottom row). In particular,  $L_{\rm BH}$ increases by almost two orders of magnitude for an one-decade decrease in $\varepsilon_{p, \rm max}^\prime$. Note also that for low values of $\varepsilon_{p, \rm max}^\prime$, the  injection luminosity of secondaries is dominated by the BH process.
As we show later in detail (Sec.~\ref{sec:4-2}), very high $L_{\rm BH}$ are disfavoured, for they result in bright electromagnetic cascade emission which is strongly constrained by the data. 
Interestingly, nearly all derived quantities are  weakly dependent on $R_b^\prime$, which means that the blob size cannot be constrained from the neutrino data alone.

By combining all the different pieces of information from Figure~\ref{fig:summary_statistic_30_2d_source_ProtonHe_b}, we consider next three indicative models (marked with star symbols in Figure~\ref{fig:summary_statistic_30_2d_source_ProtonHe_b}) that are consistent with the 2014--2015 neutrino flare observations, but differ in their predictions about the cascade emission. The model parameters are summarized in Table~\ref{tab:tab1}. In what follows, we present the results from these specific models, and discuss whether they can explain the neutrino data and blazar SED simultaneously.

\subsection{\label{sec:4-2}Indicative models for the multi-messenger emission of 2014--2015}
\begin{figure*}
\includegraphics[width=\textwidth]{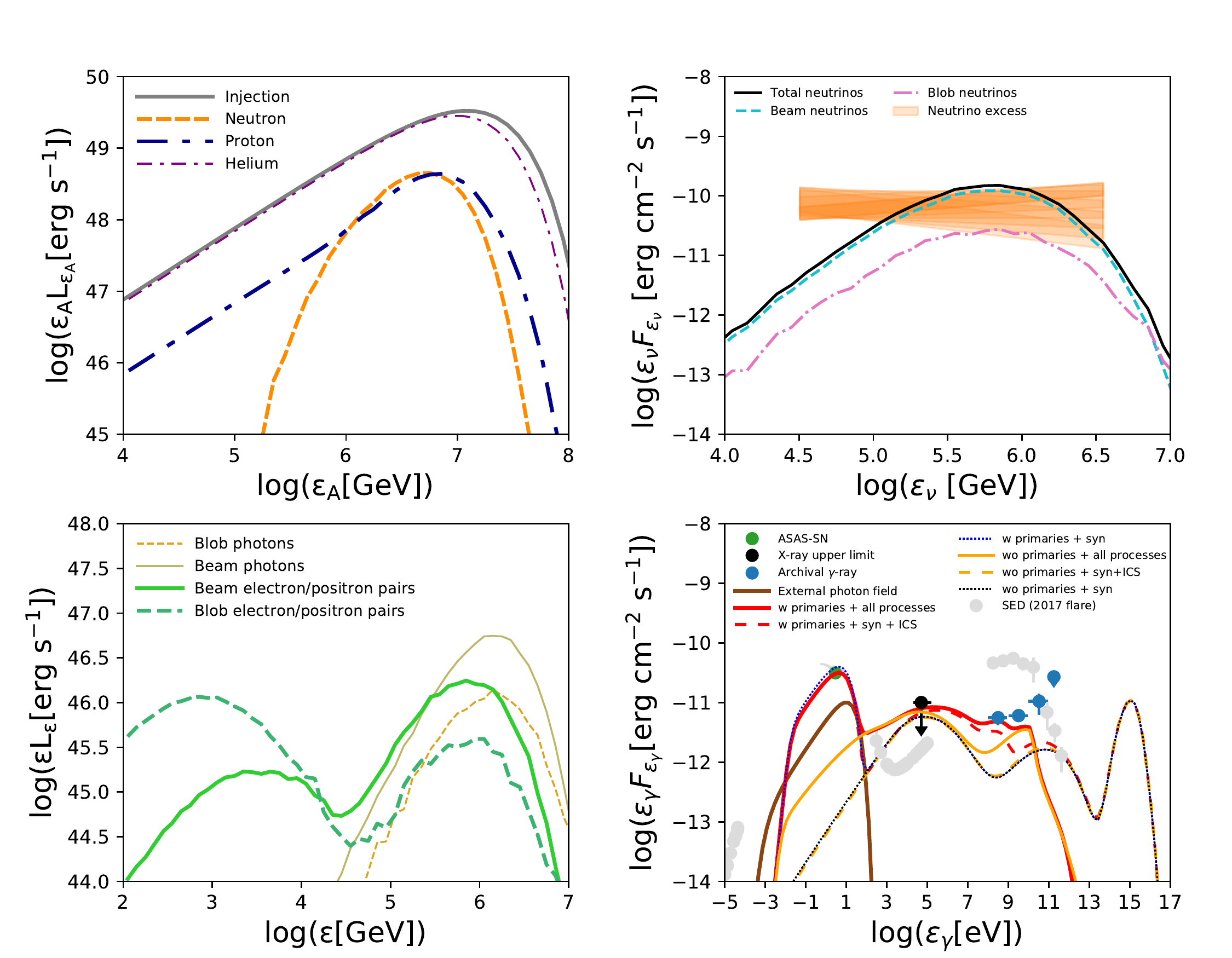}
\caption{Summary plot with results of Model A (see Table~\ref{tab:tab1}) for the 2014--2015 neutrino excess of \txs (see red star in Figure~\ref{fig:summary_statistic_30_2d_source_ProtonHe_b}). 
\textbf{Top left}: The energy spectra of CR nuclei in the blob co-moving frame obtained after one dynamical timescale (colored lines). The spectrum of injected nuclei is over plotted for comparison (solid grey line).
\textbf{Top right}: The energy spectrum of the all-flavor neutrino flux (solid black line), including the contributions of the beam  (dashed green line) and the blob (dotted-dashed cyan line). The bow tie shows the best-fit all-flavor spectrum (with its 95\% uncertainty region) obtained by IceCube \citep{IceCube:2018cha}.
\textbf{Bottom left}: The isotropic-equivalent injection luminosity of BH pairs (green lines) and $\gamma$-ray photons (orange lines) in the observer's frame. The contributions of the blob and the beam are highlighted with dashed and solid lines, respectively.  \textbf{Bottom right}: The 2014--2015 SED of \txs comprised of: optical V-band data (corrected for extinction) from ASAS-SN~\citep{Shappee2014, Kochanek2017}, hard X-ray upper limit by \textit{Swift}-BAT~\citep{Reimer:2018vvw}, and $\gamma$-ray data from \textit{Fermi}-LAT~\citep{Reimer:2018vvw}. For comparison, the SED during the 2017 flare are also shown (grey symbols)~\citep{IceCube:2018dnn}. 
Solid red and solid orange lines show the predicted SEDs with and without, respectively, the contribution of primary electrons, for $B^\prime=80$~G when all leptonic processes are considered (i.e., synchrotron radiation, ICS, and $\gamma \gamma$ pair production). For comparison, we also show the SEDs computed when only synchrotron (dotted lines) or synchrotron and ICS processes (dashed lines) are taken into account. The SED of the external radiation field is also shown (solid brown line).
\label{fig:Bestfit_2014_15_source_ProtonHe_b_Epmax_1E14_8}}
\end{figure*}

In Figure~\ref{fig:Bestfit_2014_15_source_ProtonHe_b_Epmax_1E14_8}, we show the results of Model A. In the upper left panel, we plot energy spectra of injected nuclei (gray lines) and the energy spectrum of neutrons (orange line) that are mainly produced by the photodisintegration of  helium.  
In the observer's rest frame, the total isotropic-equivalent CR injection luminosity is $L_{\rm CR} =\delta^4 L_{\rm CR}^\prime \approx 10^{50}\, {\rm~erg~s^{-1}} (\delta/10)^4 \, (L_{\rm CR}^\prime/ 10^{46}\rm~erg~s^{-1})$. We discuss energetics requirements of the model in Section~\ref{sec:5}. 

In the upper right panel, we present the all-flavor neutrino flux (black solid line) predicted by the model, with the contributions of the blob and the beam plotted with dashed-dotted and dashed lines, respectively. The neutrino flux is dominated by the beam-induced neutrinos, with the peak flux ratio of beam-induced neutrinos to blob-induced neutrinos being $\xi_\nu \sim 4$,  
in agreement with the qualitative analysis in Section~\ref{sec:2} (see also top left panel in Figure~\ref{fig:summary_statistic_30_2d_source_ProtonHe_b}). The bow-tie colored region represents the  best-fit result for the neutrino spectrum, with the 95\% statistical uncertainty on the parameter estimates, measured by IceCube with the time-dependent analysis \citep[see Figure~3 in][]{IceCube:2018cha}. The all-flavor neutrino spectrum of the model peaks at $\sim 400$~TeV (see also right panel in middle row of Figure~\ref{fig:summary_statistic_30_2d_source_ProtonHe_b}) with a flux $\varepsilon_\nu^{\rm peak} F_{\varepsilon_\nu^{\rm peak}} \simeq 1.5 \times 10^{-10}\rm~erg~cm^{-2}~s^{-1}$, as expected (see Section~\ref{sec:4-1}). Using the model-predicted muon neutrino flux and equation (\ref{eq:Nnu}) we find $\mathcal{N}_{\nu_{\mu}+\bar{\nu}_\mu} \sim 6$, consistent with the $\sim 2 \sigma$ statistical error of the IceCube results of $13\pm5$ signal muon events~\citep{IceCube:2018cha}.

In the lower left panel of Figure~\ref{fig:Bestfit_2014_15_source_ProtonHe_b_Epmax_1E14_8}, we show energy spectra of pairs (green lines) and $\gamma$-rays (orange  lines) which are produced by photo-hadronic processes\footnote{The displayed $\gamma$-ray flux does not include contributions from the de-excitation of photo-disintegrated He nuclei. The de-excitation efficiency can be estimated as $f_{\rm deex} \sim \kappa_{\rm deex} f_{A\gamma} \sim 10^{-3} f_{A\gamma}$, where $\kappa_{\rm deex} \sim 10^{-4}(56/A)$ is the energy fraction taken by $\gamma$ rays ~\citep{Murase:2010va}. The contribution of de-excitation photons from He is expected to be sub-dominant.} in the blob (dashed lines) and along the beam (solid lines). The low-energy bump of the dashed green curve corresponds to the pairs produced from the BH process of helium nuclei, whereas the second bump at higher energies is related to pairs produced by the pion decays.
Although neutrons do not pair-produce on photons, the protons produced via $n\gamma$ interactions can still produce pairs via the BH process, which leads to the low-energy hump of the solid green curve. However, the BH pair-production along the beam is a sub-dominant process for pair injection, as suggested by the peak luminosities of the two bumps in the pair injection spectrum (solid green curves). The secondaries that are injected into the blob are used for computing the electromagnetic emission of the induced cascade in that region. For reasons explained in Section \ref{sec:3}, we do not compute the emission from the beam-induced secondaries.  

In order to calculate the cascade emission in the blob, in addition to the injection rates of secondary pairs and photons (see dashed lines in the bottom left panel of Figure~\ref{fig:Bestfit_2014_15_source_ProtonHe_b_Epmax_1E14_8}), we need a value of the magnetic field strength.  
We adopt $B^\prime = 80\rm~G$ that corresponds to $u_B^\prime \sim 2 \, u_{\rm ex}^\prime$. The cascade emission from the secondaries alone is plotted with a dashed red line in the lower right panel of Figure~\ref{fig:Bestfit_2014_15_source_ProtonHe_b_Epmax_1E14_8}. The synchrotron bump of the cascade spectrum peaks at $\sim 10$ keV, with a peak flux almost saturating the {\sl Swift}-BAT upper limit. Meanwhile, the synchrotron spectrum extends to lower energies as $F_{\varepsilon}\propto \varepsilon^{-1/2}$, indicative of fast-cooling electrons. The Compton component of the cascade emission, which emerges in the {\sl Fermi}-LAT band, cannot account for the observed $\gamma$-ray flux. If we also consider the emission from a primary electron population injected into the blob (for parameters, see Table~\ref{tab:tab1}), we can explain the optical measurements of ASAS-SN (green symbol) and enhance the $\gamma$-ray flux with the ICS emission from the primaries (solid red line). However, the model has difficulty in  explaining the highest energy data point of {\sl Fermi}-LAT (i.e., possible hardening of the spectrum), due to the high $\gamma \gamma$ opacity  above $\sim 10$ GeV. 

We also explored a case with $B^\prime = 5\rm~G$ or equivalently  $u_B^\prime \ll u_{\rm ex}^\prime$ (not shown in the figure). Pairs, in this case, are cooling more efficiently via ICS. As a result, the peak of the cascade spectrum emerges in $\gamma$-rays, while the flux of the synchrotron bump (peaking at $\sim 1$ keV now due to the lower magnetic field) is suppressed by almost one order of magnitude. By adding the emission of primary electrons in this model, we can explain the {\sl Fermi} data, but the model falls short in explaining the ASAS-SN optical measurement. Any attempt to increase further the luminosity of primary electrons leads to an overshoot of the $\gamma$-ray flux due to a brighter Compton emission.

\begin{figure*}
\includegraphics[width=\textwidth]{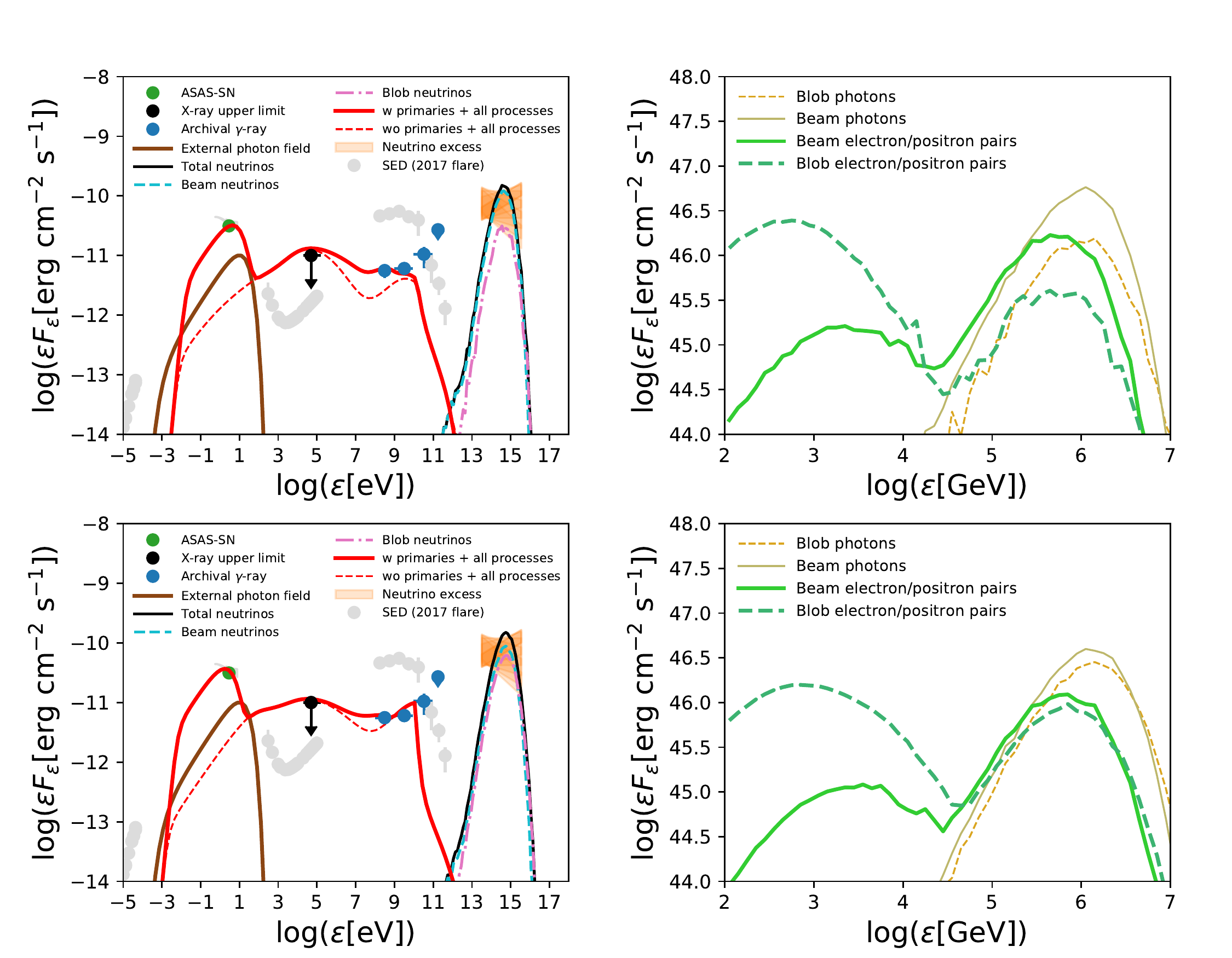}
\caption{Same as Figure~\ref{fig:Bestfit_2014_15_source_ProtonHe_b_Epmax_1E14_8}, except for models B (top panels) and C (bottom panels). The all-flavor neutrino flux and SED of \txs are shown in the left-hand side panels of the figure, while the injected (observed) luminosity of BH pairs and $\gamma$-rays is shown on the right-hand side. For the parameter values used here, see Table~\ref{tab:tab1} and Figure ~\ref{fig:summary_statistic_30_2d_source_ProtonHe_b}. 
\label{fig:Bestfit_2014_15_source_ProtonHe_b_combine}}
\end{figure*}

For comparison, we also show the results from Model B and Model C in Figure~\ref{fig:Bestfit_2014_15_source_ProtonHe_b_combine}. In Model B, we adopt a lower maximum proton energy, namely $\varepsilon_{p, \rm max}^\prime = 0.4\rm~PeV$. We find that the cascade emission alone overshoots the {\sl Swift}-BAT upper limit, because the injection luminosity of secondary electron-positron pairs is several times higher than in Model A, as shown in Figure~\ref{fig:summary_statistic_30_2d_source_ProtonHe_b}. In Model C, we consider a larger blob with radius $R_b^\prime = 6 \times 10^{15}\rm~cm$, but the same maximum proton energy as in Model A. The cascade emission is brighter than the one found for Model A, but still consistent with the {\sl Swift}-BAT upper limit. 

\subsection{\label{sec:4-2}Physical connection to the 2017 flare associated with IceCube-170922A}
In 2017, a high-energy ($E_\nu > 290\rm~TeV$) muon-track neutrino event (IceCube-170922A) was detected by IceCube's real-time alert system from the direction of \txs during a period of multi-wavelength flaring activity  \citep{IceCube:2018dnn}. The neutrino flux inferred from the detection of only one neutrino event is uncertain. For example, assuming that the neutrino emission lasted  for 0.5 years (7.5 years) the all-flavor upper limits read $\sim 1.8 \times 10^{-10} \, \rm erg~cm^{-2}~s^{-1}$ ($\sim 1.2 \times 10^{-11} \, \rm erg~cm^{-2}~s^{-1}$)  \citep{IceCube:2018dnn}. Moreover, the point-source analysis method has shown that the all-flavor neutrino flux upper limit can be one order of magnitude lower, i.e., $\sim 10^{-11} \rm erg~cm^{-2}~s^{-1}$~\citep{Keivani:2018rnh,Gao:2019NatAs}.

\begin{figure*}
\includegraphics[width=\textwidth]{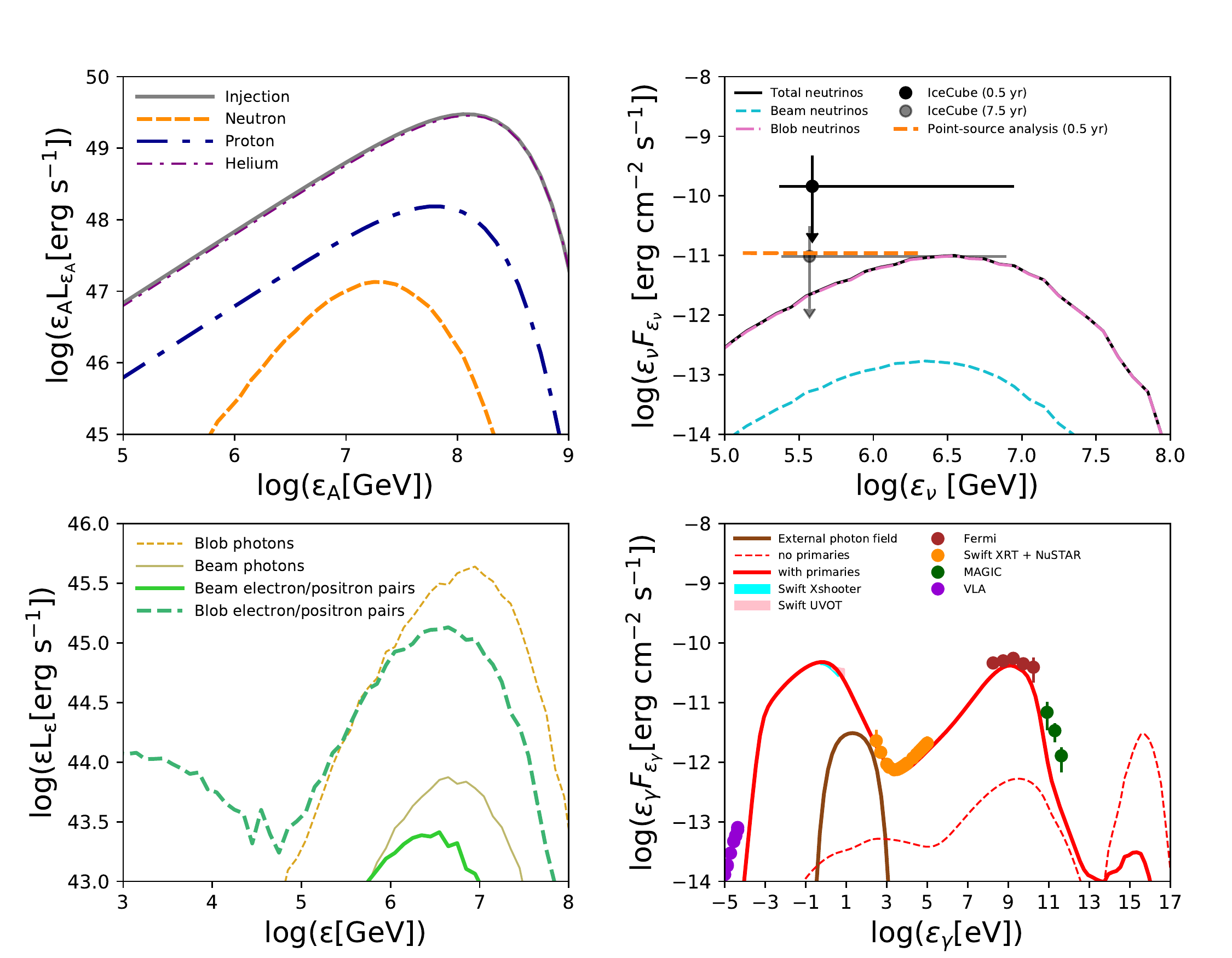}
\caption{Similar to Figure~\ref{fig:Bestfit_2014_15_source_ProtonHe_b_Epmax_1E14_8}, but for the 2017 flare with following parameters: $\varepsilon_{p, \rm max}^\prime = 10^{15.4}\rm~eV$, $R_{b}^\prime = 10^{17}\rm~cm$, and $L^\prime_{\rm CR}\simeq 1.5 \times 10^{49}$~erg s$^{-1}$. The external photon field (brown line in the lower right panel) and blob properties used here are taken from the hybrid leptonic model LMPL2b in~\cite{Keivani:2018rnh}, $\varepsilon_{\rm min}^\prime = 0.5\rm~eV$, $\varepsilon_{\rm max}^\prime = 5\rm~keV$, $s_{\rm ex} = 2$, $u_{\rm ex}^\prime = 0.08\rm~erg~cm^{-3}$, $R_{\rm ex} = 2 \times 10^{19}\rm~cm$, and $\delta = 25$ (for a complete list of parameters, see Table~\ref{tab:tab2}). Note that the very-high-energy $\gamma$-ray data measured by MAGIC were not included in the SED fitting of \citet{Keivani:2018rnh}, as they were not publicly available at the time. 
\label{fig:summary_neutrino_AGN_ExPowerlaw35_2017_source_ProtonHe_b}}
\end{figure*}

Here, we do not aim to explain the SED of \txs in 2017 from scratch. Instead, we adopt similar parameter values for the blob and external photon field as those used in  \cite{Keivani:2018rnh} to explain the SED. Note that all models presented by \cite{Keivani:2018rnh} considered a larger blob and external photon fields with much lower energy density than those used here for the 2014--2015 period. The questions we want to address here are the following: 
\begin{itemize}
    \item Are the predictions of the neutral beam model~\citep{Murase:2018iyl} quantitatively consistent with the detection of \icnu, if protons and helium nuclei were injected into a less compact blob as found by \cite{Keivani:2018rnh}?
    \item What would be the contribution of the neutral beam to the total neutrino emission of \txs in 2017? 
\end{itemize}
We adopt similar parameters as those used in the model LMPL2b of \cite{Keivani:2018rnh}:
$u_{\rm ex}^\prime = 0.08\rm~erg~cm^{-3}$, $\varepsilon_{\rm min}^\prime = 50 \rm~eV$, $\varepsilon_{\rm max}^\prime = 5\rm~keV$, $s_{\rm ex} = 2$, $R_{\rm ex} =10^{19}\rm~cm$, $R_b^\prime = 10^{17}\rm~cm$ (for a detailed list of input model parameters, see Table~\ref{tab:tab2}).
Our results for $\varepsilon_{p, \rm max}^\prime = 10^{15.4}\rm~eV$, $R_{b}^\prime = 10^{17}\rm~cm$, and $L_{\rm CR}= 8.2 \times 10^{49}$~erg s$^{-1}$ are summarized in Figure~\ref{fig:summary_neutrino_AGN_ExPowerlaw35_2017_source_ProtonHe_b}.  We find that the neutral beam model, when applied to the 2017 flare, yields results that are consistent with the detection of $\sim 1$ muon neutrino event, even though the emission of the beam-induced neutrinos is $\sim 100$ times lower than that of blob-induced neutrinos.

If we lower the value of the minimum photon energy (without changing any other parameter) down to, e.g., $\varepsilon_{\rm min}^\prime = 0.5 \rm~eV$, then the emission of the beam-induced neutrinos becomes comparable to that from the blob. In that case, the total neutrino flux is $\sim 2$ times higher than the flux predicted by standard single-zone models that consider only the neutrino emission from the blob.
However, we cannot readily simultaneously explain well the observed SED of the 2017 flare for the following reason: by decreasing $\varepsilon^\prime_{\min}$ we do not only increase the interaction efficiency of helium nuclei to produce secondaries, but we also enhance the inverse Compton scattering rate between electrons and lower energy photons. The latter leads to an enhancement of the SSC emission, which overshoots the {\sl Swift}/XRT and {\sl NuSTAR} data. 

The properties of the external photon field needed to explain both the neutrino flux and the SED of the 2017 flare are coupled to the parameters describing the blob and the relativistic particles therein. Here, we chose similar parameters as those used in \citet{Keivani:2018rnh}. It is therefore likely that other combinations of parameters, which can explain the SED, may at the same time allow for a higher contribution of beam-induced neutrinos to the neutrino flux., by e.g., allowing the use of larger $R_{\rm ex}$~\citep{Murase:2018iyl} and/or higher $u^\prime_{\rm ex}$ and/or lower $\epsilon^\prime_{\min}$ than those adopted here. A wide parameter space search for the 2017 flare lies, however, beyond the scope of this work. 

\begin{table} 
\caption{\label{tab:tab2}A list of input model parameters for the 2017 flare.}
\begin{ruledtabular}
\begin{tabular}{llll}
\textrm{Physical parameters} & \multicolumn{3}{c}{\textrm{Value}}\\
\colrule
\rule{0pt}{3ex} \\
\multicolumn{4}{l}{\textbf{Default parameters used in this work}}  \\
External photon field radius ($R_{\rm ex}$ [cm]) & \multicolumn{3}{c}{$10^{19}$} \\
External energy density ($u_{\rm ex}^\prime$ [erg/cm$^{3}$]) & \multicolumn{3}{c}{$0.08$} \\
External photon spectral index ($s_{\rm ex}$) & \multicolumn{3}{c}{2} \\
Minimum photon energy ($\varepsilon_{\rm min}^\prime$ [keV]) & \multicolumn{3}{c}{$0.5$} \\
Maximum photon energy ($\varepsilon_{\rm max}^\prime$ [keV]) & \multicolumn{3}{c}{$5$} \\
Blob Lorentz factor ($\Gamma$) & \multicolumn{3}{c}{$25$} \\
Minimum proton energy ($\varepsilon_{p, \rm max}^\prime$ [GeV]) & \multicolumn{3}{c}{$1$}\\
Nuclei acceleration spectral index ($s_{\rm acc}$) & \multicolumn{3}{c}{1} \\
Number ratio of accelerated nuclei ($f_{\rm He} / f_{\rm P}$) & \multicolumn{3}{c}{5} \\
\rule{0pt}{3ex} \\
\multicolumn{4}{l}{\textbf{Optimized parameters for  neutrino flux}} \\ 
Maximum proton energy ($\varepsilon_{p, \rm max}^\prime$ [PeV]) & &2.5& \\
Blob radius ($R_b^\prime$ [$10^{15}$ cm]) & &$100$ & \\
Total CR luminosity ($L_{\rm CR}$ [$10^{49}$ erg/s]) & &8 &\\
\rule{0pt}{3ex} \\
\multicolumn{4}{l}{\textbf{Additional parameters for SED}} \\
Magnetic field strength ($B^\prime$ [G]) & & 0.4 & \\
Minimum electron energy ($\varepsilon_{e, \rm min}^\prime$ [MeV]) & & 0.5&  \\
Break electron energy ($\varepsilon_{e, \rm br}^\prime$ [GeV]) & & 2&  \\
Maximum electron energy ($\varepsilon_{e, \rm max}^\prime$ [GeV]) & & 40 &  \\
Spectral index before break ($s_{\rm acc, 1}$) &  &  1.9 \\
Spectral index above break ($s_{\rm acc, 2}$) &  &  3.6 \\
Total electron luminosity ($L_{\rm e}$ [$10^{46}$ erg/s]) & & 74& \\
\end{tabular}
\end{ruledtabular}
\tablecomments{The parameter values used for the external photon field and the blob  are similar to those used in \citet{Keivani:2018rnh} (see LMPL2b model in Tables~6 and 7 therein).}
\end{table}

\section{\label{sec:5} Summary and Discussion}
The neutral beam model has been suggested to explain flaring neutrino emissions of \txs~\citep{Murase:2018iyl}. We presented the first comprehensive study for the neutrino emission from \txs in the framework of the neutral beam model for blazars. We demonstrated that both the 2014--2015 neutrino flare excess and the 2017 multi-messenger flare can be explained by the neutral beam model without violating the X-ray and $\gamma$-ray observations. 

Our Model A for the 2014--2015 neutrino excess predicts a number of $\mathcal{N}_{\nu_\mu + \bar{\nu}_\mu} \sim 6$ muon-track neutrino events within a period of 158 days and the energy range of 30 TeV to 2 PeV, consistent with the IceCube detection within the $2\sigma$ uncertainty range. The total all-flavor  neutrino flux at the peak neutrino energy $\varepsilon_{\nu, \rm peak}\simeq 400\rm~TeV$ is $\varepsilon_\nu^{\rm peak} F_{{\varepsilon_\nu}^{\rm peak}} \simeq 1.5 \times 10^{-10}\rm~erg~cm^{-2}~s^{-1}$, with the contribution of beam-induced neutrinos being $\sim 4$ 
times larger than the contribution of blob-induced neutrinos. 
The electromagnetic cascade emission of the default model is consistent with multi-wavelength data and X-ray upper limits. By also considering the emission of primary electrons in the blob, we showed that the observed SED can be adequately described if $B^\prime = 80\rm~G$. Lower values of the magnetic field (i.e., $B^\prime \ll \sqrt{8\pi u^\prime_{\rm ext}}$) are disfavored, for they lead to bright inverse Compton emission that overshoots the {\sl Fermi}-LAT data. 
Small changes in either the blob radius $R^\prime_b$ or maximum proton energy $\varepsilon_{p,\rm max}^\prime$ compared to their default values (i.e., $10^{15}\rm~cm$ and $10^{14.8}\rm~eV$, respectively) can enhance the cascade emission in the blob, especially in the latter case.

We also showed that the neutral beam model, when applied to the 2017 flare, results in a neutrino flux that is dominated by the blob emission and is consistent with the upper limits inferred from the detection of \icnu. The neutral beam model therefore provides a common physical framework for explaining the IceCube observations of both epochs. This can be achieved, however, only if the properties of the blob are significantly  different between 2014--2015 and 2017. More specifically, we find that the blob should be more compact, with stronger magnetic field, and lower Lorentz factor in 2014--2015 compared to 2017. These results also suggest that the blob was formed closer to the central black during the period of the neutrino excess compared to 2017, where a larger dissipation distance is more plausible (e.g., $z \approx R^\prime_b / \theta_j \simeq 10^{18}~{\rm cm}\, (R^\prime_b/10^{17}~{\rm cm})(0.1/\theta_j)$, for a conical jet with opening angle $\theta_j$). In addition to the blob properties, which differ significantly between the two epochs, the CR co-moving injection luminosity also differs by a factor of $\sim 40$, with higher luminosities required for the period of the neutrino excess.

Here, we focused on two epochs of interest for \txs \, from the entire IceCube lifetime (9.5 years), and showed that the neutral beam model provides a common physical framework for both. Then, the question about the model's predictions for the blazar's long-term neutrino emission naturally arises. In the context of the neutral beam model, the non-detection of neutrino fluxes as high as that of the 2014--2015 excess during the IceCube lifetime, implies that the dissipation that led to the neutrino excess is not continuous. In addition, the conditions necessary to explain the neutrino excess (i.e., compact dissipation region with strong magnetic fields, and high-density UV/soft X-ray radiation field), point to dissipation occurring close to the SMBH. An interesting possibility is the interaction of the blazar's jet with misaligned sub-disks, as recently demonstrated with general relativistic magnetohydrodynamic simulations of tilted thin accretion disks \citep{Liska2019}. The jet-disk interaction can induce magnetic instabilities and current sheets, leading to energy dissipation and CR acceleration (A.~Tchekhovskoy, private communication), while the sub-disks can provide dense radiation fields for photohadronic and photodisintegration processes of nuclei, as well as for attenuation of very high energy $\gamma$-rays. Nevertheless, continuous dissipation occurring within the jet and at large distances from the SMBH, as inferred from our modeling of the 2017 flare, is still possible. In this case, the dissipation region can be associated with the so-called blazar zone, where the bulk of the blazar's emission is produced. \cite{Petropoulou2019arXiv} studied the long-term neutrino emission of \txs \, from the blazar zone, assuming that the latter has the same properties as our model's blob in 2017. These authors derived a conservative estimate of $\sim 0.4-2$ muon neutrinos over a $\sim10$-year long period of IceCube observations, which is consistent with the detection of \icnu \, being an upper-fluctuation instead of being really associated with the 2017 flare.
 
In this work, we considered an arbitrary external photon field as the main target for the photo-hadronic interactions in  the blob and the beam. For the modeling of the 2014--2015 neutrino excess, a very dense external photon field is required, with co-moving energy density $u_{\rm ex}^\prime \sim 100~{\rm erg}~{\rm cm}^{-3}$ (see Table~\ref{tab:tab1}). The latter is similar to the value  
found in~\cite{Reimer:2018vvw}, who searched for the minimal target photon field needed to produce the neutrino emission of 2014--2015.
However, we cannot explain the SED of the 2017 flare with such dense external photon field. \citet{Keivani:2018rnh} showed that the SED of the 2017 flare can be well-modeled with a much lower energy density of external photon fields (i.e., $u_{\rm ex}^\prime \lesssim 0.1~{\rm erg}~{\rm cm}^{-3}$).
The observed luminosity of the external photon field used here to explain the 2014--2015 flare is $L_{\rm ex} \sim 10^{46}\rm~erg~s^{-1}$. This is also similar to the value inferred by the modeling of the 2017 flare (see Table~\ref{tab:tab2} and \citet{Keivani:2018rnh}), which would not be so far from the maximally allowed luminosity based on the SED during the 2017 flare.

It is still an open question how a similar external radiation luminosity is realized in the 2014--2015 and 2017 epochs, while the density and size of the external radiation field as well as the properties of the dissipation region are significantly different. The luminosity of the external photon field is extreme when compared to the typical luminosity of the emission from broad-line region, $L_{\rm BLR} \sim 5 \times 10^{43}\rm~erg~s^{-1}$, or accretion disk, $L_{\rm AD} \sim 3 \times 10^{44}\rm~erg~s^{-1}$~\citep{Padovani:2019xcv}. 
The broadband external photon field could arise from the sheath region of a structured jet, where CRs accelerated in a faster spine interact with photons emitted by electrons accelerated in the slower sheath region~\citep{Ghisellini:2004ec,Tavecchio:2014iza,Righi:2016kio,Ahnen:2018mvi,Tavecchio:2019nvg}. We note, however, that $L_{\rm ex}\sim 10^{46}$~erg s$^{-1}$ is much higher than the luminosities of the sheath typically inferred by the modeling of other blazars \citep[e.g.,][]{Ghisellini:2004ec}. Variable external radiation fields are naturally expected in this scenario, although the details (e.g., timescale of variations) remain unclear. Interestingly, the presence of a variable external photon field on month-long timescales is also inferred by the SED modeling of archival data (Petropoulou et al. in preparation), even though the necessary changes in the photon energy density as not as extreme as found here (i.e., from $\sim 100\rm~erg~cm^{-3}$ down to $\sim 0.3\rm~erg~cm^{-3}$). \cite{2019A&A...630A.103B} argued that \txs is a special blazar, hosting a SMBH binary that produces two interacting precessing jets. If this scenario is confirmed, one of the two jets could in principle provide an additional external photon field. However, given the large physical distance between the two jets, $R_{\rm ex}\sim10^{19}$~cm, \citep{2019A&A...630A.103B}, the photon energy density in one jet (with an observed luminosity $L_{\rm ex}\lesssim 10^{46}$~erg s$^{-1}$) as seen in the rest frame of the other one is typically small, $\lesssim{\rm a~few}$~erg cm$^{-3}$, even when the Doppler boosting due to the relative motion of the two jets is taken into account. Such photon energy densities seem insufficient to explain the 2014--2015 neutrino flare.

The absolute jet power of \txs can be written as:
\begin{equation}
P_j = \eta_j \, L_{\rm Edd} \simeq 1.5 \times 10^{47} \, \left(\frac{\eta_j}{0.9}\right) \left(\frac{M_{\rm BH}}{10^9 M_{\odot}}\right) \rm~erg~s^{-1},
\end{equation}
where $L_{\rm Edd}\simeq 1.7 \times 10^{47}\, (M_{\rm BH}/10^9 M_{\odot})$~erg s$^{-1}$ is the Eddington luminosity of a SMBH with mass $M_{\rm BH}$ and $\eta_j\le 1$ is an efficiency factor. If a fraction $\epsilon_{\rm CR}$ of the jet power is carried by relativistic protons and nuclei, then we can estimate the isotropic-equivalent CR luminosity as~\citep{Murase:2018iyl} 
\begin{eqnarray}
L_{\rm CR}& \approx & \frac{2}{\theta_{\rm beam}^2} \frac{b_{\rm fl}^\nu}{f_{\rm fl}^\nu}\epsilon_{\rm CR} P_j\nonumber \\
&\sim &10^{50}  \left(\frac{\theta_{\rm beam}}{0.1}\right)^{-2} \left(\frac{b_{\rm fl}^\nu/f_{\rm fl}^\nu}{10}\right) \left(\frac{\epsilon_{\rm CR}}{0.3} \right) \nonumber\\ 
&\times& (\eta_j/0.9) (M_{\rm BH}/10^9 M_{\odot}) \, {\rm erg \, s}^{-1},
\label{eq:Lcr}
\end{eqnarray}
where $\theta_{\rm beam} \sim 1/\Gamma \sim 0.1$ is the opening angle of the beam, and $b_{\rm fl}^\nu/f_{\rm fl}^\nu$ is the ratio of the energy dissipated during flares $b_{\rm fl}$ to the fraction of time spent in a flaring state $f_{\rm fl}$ and should be larger than unity for  flares~\citep{Murase:2018iyl}.
The CR luminosity derived in Model A is therefore plausible, only if $\epsilon_{\rm CR} \sim 0.3$, $\eta_j \sim 0.9$, and $ b_{\rm fl}^\nu/f_{\rm fl}^\nu \sim 10$.  Note that energetics requirements would be even more excessive, had we assumed a softer CR injection spectrum (with e.g., $s_{\rm acc}\gtrsim 2$). 

We considered the case where CRs loaded into the jet have a mixed composition that is mainly composed of protons and helium.  We explored different values for the ratio of helium to protons, ranging from 1/12 to a pure helium composition. We found that the contribution of beam-induced neutrinos increases for larger values of the $f_{\rm He}/f_{\rm P}$ ratio, as  more free neutrons are generated via the photodisintegration of helium nuclei. 
We also tried other compositions, which are dominated by heavy nuclei, but we had difficulty in finding parameters that can explain the IceCube neutrino data without overshooting the electromagnetic data. This is because the energy loss efficiency of the BH process is sensitive to the ratio of nuclei charge number to mass number,i.e., $f_{{\rm BH}, A} \propto Z^2/A$. For example, the BH pair luminosity for fully ionized iron nuclei can be $\sim 10$ times larger than those for protons or helium. 
In the case of CR injection with solar composition, the CRs loaded into the jet are mainly dominated by protons. In this case, the flux of the blob-induced neutrinos is larger than the flux of beam-induced neutrinos, because the flux of the neutral beam is suppressed due to the low efficiency of the photomeson production process.




In conclusion, the neutral beam model can provide a common framework for explaining the neutrino and electromagnetic emission of \txs in both periods of the 2014--2015 and 2017 flares if: (i) the ratio of helium to protons for accelerated CRs is about 5 or beyond, (ii) the external radiation field is strong enough in both of the flare cases and may vary in month-to-year timescales, (iii) the injection CR luminosity, the Lorentz factor, and comoving size of the blob also vary in month-to-year timescales, and (iv) electromagnetic cascades induced by the neutron beam are developed in the decelerated jet or interstellar medium. 

\acknowledgments
The authors thank the anonymous referee for their constructive report. B.T.Z. is supported by High-performance Computing Platform of Peking University.
M.P. acknowledges support from the Lyman Spitzer, Jr.~Postdoctoral Fellowship and from the Fermi Guest Investigation grant 80NSSC18K1745.
The work of K.M. is supported by NSF Grant No.~PHY-1620777 and AST-1908689, and the Alfred P. Sloan Foundation. F.O. acknowledges support by the Deutsche Forschungsgemeinschaft through grant 
SFB\,1258 ``Neutrinos and Dark Matter in Astro- and Particle Physics''.

\software{CRPropa (v3.0; Alves Batista et al. 2016)}
\bibliography{bzhang}
\end{document}